\documentclass[%
 reprint,
 superscriptaddress,
 amsmath,amssymb,
 aps,
prx,
floatfix,
]{revtex4-2}
\usepackage{graphicx}
\usepackage{dcolumn}
\usepackage{bm}
\usepackage{siunitx}
\usepackage[columnwise]{lineno}
\usepackage{algorithm}[compatible]
\usepackage[noend]{algpseudocode} 
\usepackage[export]{adjustbox}
\usepackage{rotating}

\usepackage{color}
\usepackage[usenames, dvipsnames]{xcolor}

\usepackage{hyperref}

\begin{document}
\preprint{}
\title{Tuning Arrays with Rays: Physics-Informed Tuning of Quantum Dot Charge States}

\author{Joshua Ziegler}
\affiliation{National Institute of Standards and Technology, Gaithersburg, Maryland 20899, USA}

\author{Florian Luthi}
\affiliation{Intel Components Research, Intel Corporation, 2501 NW 229th Avenue, Hillsboro, Oregon 97124, USA}

\author{Mick Ramsey}
\affiliation{Intel Components Research, Intel Corporation, 2501 NW 229th Avenue, Hillsboro, Oregon 97124, USA}

\author{Felix Borjans}
\affiliation{Intel Components Research, Intel Corporation, 2501 NW 229th Avenue, Hillsboro, Oregon 97124, USA}

\author{Guoji Zheng}
\affiliation{Intel Components Research, Intel Corporation, 2501 NW 229th Avenue, Hillsboro, Oregon 97124, USA}

\author{Justyna P. Zwolak}
\email{jpzwolak@nist.gov}
\affiliation{National Institute of Standards and Technology, Gaithersburg, Maryland 20899, USA}
\affiliation{Joint Center for Quantum Information and Computer Science, University of Maryland, College Park, Maryland 20742, USA}

\date{\today}
\begin{abstract}
Quantum computers based on gate-defined quantum dots (QDs) are expected to scale. 
However, as the number of qubits increases, the burden of manually calibrating these systems becomes unreasonable and autonomous tuning must be used. 
There has been a range of recent demonstrations of automated tuning of various QD parameters such as coarse gate ranges, global state topology (e.g. single QD, double QD), charge, and tunnel coupling with a variety of methods. 
Here, we demonstrate an intuitive, reliable, and data-efficient set of tools for an automated global state and charge tuning in a framework deemed \emph{physics-informed tuning} (PIT). 
The first module of PIT is an \emph{action-based} algorithm that combines a machine learning classifier with physics knowledge to navigate to a target global state. 
The second module uses a series of one-dimensional measurements to tune to a target charge state by first emptying the QDs of charge, followed by calibrating capacitive couplings and navigating to the target charge state. 
The success rate for the action-based tuning consistently surpasses $95~\%$ on both simulated and experimental data suitable for off-line testing. 
The success rate for charge setting is comparable when testing with simulated data, at $95.5(5.4)~\%$, and only slightly worse for off-line experimental tests, with an average of $89.7(17.4)~\%$ (median $97.5~\%$). 
It is noteworthy that the high performance is demonstrated both on data from samples fabricated in an academic cleanroom as well as on an industrial 300-\si{\milli\meter} process line, further underlining the device agnosticism of PIT.
Together, these tests on a range of simulated and experimental devices demonstrate the effectiveness and robustness of PIT.
\end{abstract}

\maketitle
\section{Introduction}
Quantum dot (QD) arrays, in which individual charge carriers are trapped in localized potential wells, are a promising platform to realize useful quantum computing applications~\cite{Vandersypen17-ISQ, Li18-CNQ, Boter22-SWA}.
Advantages of this platform include a small device footprint~\cite{Ansaloni20-FFQ, Pillarisetty21-SQC, Zwerver22-QMA, Weinstein22-ULS}, compatibility with industrial semiconductor fabrication techniques~\cite{Ansaloni20-FFQ, Pillarisetty21-SQC, Zwerver22-QMA}, and potential for operation with baseband pulses~\cite{Weinstein22-ULS}.
However, because single charge carriers have electrochemical sensitivity to minor impurities or imperfections, calibration and tuning of QD devices is a nontrivial and time-consuming process. 
Each QD requires a careful adjustment of gate voltages to define charge number and tunnel couplings to other QDs or reservoirs~\cite{McJunkin21-PhD, Volk19-LQR}.
Although manual calibration is achievable for small, few-QD devices, the control parameter space grows quickly for larger arrays, necessitating autonomous tune-up.
Moreover, even for automated methods, the size of the voltage gate space that must be explored can become prohibitively large.
Reducing the dimensionality of the voltage space for large QD arrays to a series of single- and double-QD systems will be important for mitigating this challenge~\cite{Oosterkamp98-MSQ, Hensgens17-FHQ, Volk19-LQR}.
As device integration improves and moves from few-QD to many-QD devices, autotuning algorithms must maximize the information of measured data to ensure efficient tuning.

To date, there have been numerous demonstrations of automation for the various phases of the tuning process for both single- and double-QD devices~\cite{Zwolak21-AAQ}.
Some approaches seek to tackle tuning starting from device turn-on to coarse tuning~\cite{Baart16-CAT, Darulova19-ATQ, Moon20-ATQ, Czischek21-MNA}.
Other methods assume that bootstrapping and basic tuning have been completed, leading to more targeted automation and coarse-tuning approaches~\cite{Zwolak18-QLD, Durrer19-ATQ, Zwolak20-AQD, Zwolak21-RBI, Lapointe-Major19-ATQ}.
However, the imperfect machine learning (ML) predictions or suboptimal fitting routines implemented in these algorithms can cause unexpected failures. 
These problems can be largely mitigated by training and testing on suitable data~\cite{Darulova20-EDM, Ziegler22-TRA} and employing data quality control (DQC) systems~\cite{Ziegler22-TRA}.
Major factors hindering the efficiency and effectiveness of navigation of the voltage space are large voltage sweeps and simplistic navigation methods.

Here, we demonstrate a physics-informed tuning (PIT) algorithm for navigating to a target charge configuration that is both effective and data efficient~\cite{Zwolak22-Patent}.
PIT consists of a pair of modules for automated coarse and charge tuning of a double-QD system.
The coarse-tuning module combines a data quality control system and a robust ML classifier~\cite{Zwolak20-AQD, Zwolak21-RBI} with device physics knowledge and heuristics (as opposed to using classical optimization or ML) to navigate to a target global state corresponding to a set of charge islands (i.e., single or double  QD).
The charge-tuning module uses a series of one-dimensional (1D) measurements and custom peak-finding techniques to find a desired charge configuration.
The incorporation of virtual gates during charge tuning allows for intuitive, efficient, and reliable targeted loading.
Together, these modules navigate from a roughly estimated starting point to a target charge state.
In particular, for the starting point, we assume that reasonable estimates for each plunger and barrier gate voltage of the double-QD system were previously identified and that the charge sensor is calibrated~\cite{Hader23-NST}.

\begin{figure*}[t]
\centering
\includegraphics[width=\linewidth]{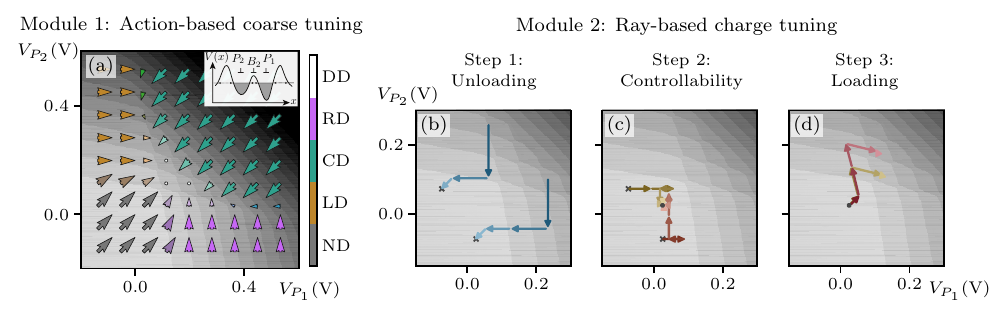}
\caption{
The flow of the PIT algorithm visualized using an idealized simulated double-QD device.
(a) The action-based coarse-tuning module combines ML state predictions with the overall QD state topology to navigate the $(P_1, P_2)$ plunger-plunger gate space.
The orientation and size of the arrows overlaying the scan correspond to the suggested gate voltage adjustment direction and magnitude, respectively.
The expected outcome for the coarse-tuning module is a gate voltage configuration defining a double-QD state.
The inset shows a representative double-QD potential profile.
(b)--(d) The ray-based charge-tuning module.
The charge-tuning process involves three steps: 
(b) unloading the double QD of all electrons using the physical gates space with the termination point marked with an ``x'',
(c) tuning to a region near the first charge transitions for both QDs (marked with a dot) and determining virtual gates,
(d) loading the desired number of electrons on each QD using the virtual gate space.
Panels (b)--(d) show charge-tuning paths for two sample points, with the magnitude of the arrows representing the size of the consecutive steps and the color lightness indicating the progress of the unloading process. 
}
\label{fig:fig-1} 
\end{figure*}

We show that this approach can be highly successful in off-line tests on both realistic simulated data~\cite{Ziegler22-TRA} and large experimentally acquired two-dimensional (2D) scans~\cite{qf-data, Pillarisetty21-SQC}.
Furthermore, by incorporating a DQC system~\cite{Ziegler22-TRA} into the action-based coarse-tuning module, and using noise-robust peak-finding tools for charge configuration, we show that the failure modes observed in off-line tests can be largely avoided.
To do so, we build an interactive simulated tuner with built-in realistic noise variation mimicking experimental conditions and device recalibration, and consistently find a charge-setting success rate around $95~\%$, regardless of the target charge configuration.
Together, these results demonstrate an automated approach for tuning a double-QD system to a target charge state in a data-efficient manner without sacrificing effectiveness---a core component for the autotuning of large QD arrays~\cite{Oosterkamp98-MSQ, Hensgens17-FHQ, Volk19-LQR, Liu22-ACT}.
Gate virtualization, also implemented in PIT, enables the isolation of the chemical potentials, allowing for targeted tuning of individual double-QD systems in a larger QD array using PIT. 
As PIT does not make strong assumptions on the device connectivity, autonomous tune-up of large 2D arrays with almost arbitrary connectivity becomes feasible.

The remainder of the paper is organized as follows: In Sec.~\ref{sec:methods} we give an overview of the design of the PIT algorithm.
The specifics of the PIT configuration for all tests in this work are described in Sec.~\ref{ssec:config}.
The performance on simulated and experimental data is discussed in Secs.~\ref{ssec:perf-sim} and \ref{ssec:perf-exp}, respectively.
Results of the interactive tuning are presented in Sec.~\ref{ssec:tuning-playground}.
We conclude with a discussion of the potential modifications to further improve the proposed autotuning technique in Sec.~\ref{sec:conclusion}.

\section{Physics-Informed Tuning: Methods}
\label{sec:methods}
The flow of the PIT algorithm is depicted in Fig.~\ref{fig:fig-1}.
PIT assumes that the device initialization (bootstrapping) is complete and that the device is brought into an appropriate parameter range for coarse tuning.
This includes establishing operational local sensing systems, determining a reference of acceptable parameter variations, i.e., the safety limits for all gates (``sandbox''), and approximating charging energies for each QD (extracted from the spacing between Coulomb peaks during pinch-off tests; for a detailed discussion of the bootstrapping phase, see, e.g., Ref.~\cite{Zwolak21-AAQ}).
The safe ranges for the plunger gates are used to determine the measurement limits and the approximate charging energies are used to determine the ray length (for all ray-based measurements) and the size of the 2D images (for the image-based coarse tuning).
We also assume that charge carriers in the device are electrons when discussing the signs of voltage changes; for holes, the signs would need to be reversed.

The PIT algorithm consists of two modules: the {\it action-based coarse tuning} and the {\it ray-based charge tuning}; see Fig.~\ref{fig:fig-1}.
The coarse-tuning module combines ML techniques~\cite{Kalantre17-MLD}---used to assess the captured global state of the device in the relevant gate space---with a physics-inspired navigation algorithm. 
The global state here means the set of charge islands formed in the device while the charge state is the exact charge configuration on the QDs. 
Depending on the experimental setup, the state assessment in PIT can be done using traditional 2D images~\cite{Kalantre17-MLD, Zwolak20-AQD} or via 1D rays~\cite{Zwolak20-RBC, Zwolak21-RBI, Chatterjee21-AEC}.

The charge-tuning module implemented in PIT follows the typical strategy of first unloading the QDs of all charges and then loading the desired number of charges on each QD~\cite{Lapointe-Major19-ATQ, Czischek21-MNA, Durrer19-ATQ}.
However, unlike in previous implementations, the emptying phase in PIT is followed by a gate-virtualization step~\cite{Hensgens17-FHQ, Hensgens18-PhD, Perron15-QSB} to ensure targeted control of each QD.
The linear combinations of gate voltages determined in this phase compensate for the capacitive cross-talk and allow for control of the electrochemical potential of individual QDs during the loading process~\cite{Oosterkamp98-MSQ}.
A high-level description of both modules is presented in the following sections.
The specific details about the PIT configuration used to perform all tests reported in this work are presented in Sec.~\ref{ssec:config}.

\subsection{Action-based coarse-tuning module}
\label{ssec:abt}
The action-based coarse-tuning algorithm used in PIT to bring the device to the desired global state leverages the overall state topology in the gate space and the intended effect of each plunger gate on the global device state.
Ideally, changing voltages on a particular plunger gate should only affect the electrochemical potential (and, thus, the number of charge carriers) on the QD it is designed to control, e.g., gate $P_2$ should load electrons onto the left dot and $P_1$ should load electrons onto the right dot; see the inset in Fig.~\ref{fig:fig-1}(a).
Because of the capacitive cross-talk between the gate electrodes, such fine control of charge occupation with individual gates is not possible; however, the relationship between globally defined states is preserved. 
It is therefore possible to move from a left-QD state towards a double QD by steadily increasing $V_{P_1}$ or from a central single QD to double QD by decreasing both plunger gates; see Fig.~\ref{fig:fig-1}(a).

The action-based tuning combines a ML algorithm used to determine the global state of the device near a given point in the $(V_{P_1}, V_{P_2})$ plunger-plunger gate voltage space with a physics-informed navigation strategy. 
It takes as an input a point $x_c\in(V_{P_1},V_{P_2})$, an acceptable exit threshold $\delta_{\rm tr}$, and a preferred classifier.
Provided the choice of a state classifier, the PIT algorithm initiates either a series of ray-based measurements [for the ray-based classification (RBC) framework~\cite{Zwolak20-RBC, Zwolak21-RBI}] or 2D scans (for a convolutional neural network-based model~\cite{Kalantre17-MLD, Zwolak20-AQD}) followed by the state assessment.
The returned state vector,
\begin{equation}\label{eq:prob_vec}
{\rm\bf{p}}(x_{\rm c})=(p_{\rm ND},\,p_{{\rm SD}_L},\,p_{{\rm SD}_C},\,p_{{\rm SD}_R},\,p_{\rm DD}),
\end{equation} 
represents the probability of each possible global state being captured in the measurement, with ND indicating that no QD is formed, ${\rm SD}_L$, ${\rm SD}_C$, and ${\rm SD}_R$ denoting the left, central, and right single QD, respectively, and ${\rm DD}$ denoting the double-QD state [see Fig.~\ref{fig:fig-1}(a)].
If ${\rm\bf{p}}(x_{\rm c})$ is sufficiently close to the target DD state, as measured by some distance function $\delta$, the algorithm terminates.
Otherwise, a subsequent measurement is initiated at a voltage configuration determined by the action vector,
\begin{equation}\label{eq:act_vect}
{\rm\bf{v}}_{\rm act}:=(V_{P_1}^{\rm act},V_{P_2}^{\rm act})={\rm\bf{p}}(x_{\rm c})\cdot\mathbb{A}^T,
\end{equation} 
where $(\cdot)^T$ is a matrix transpose and the action array is defined as
\begin{equation}\label{eq:act_array}
\mathbb{A} = \alpha\,{\rm diag}({\rm\bf{v}}_{E_C})\cdot
\begin{pmatrix}
1 & 1 & -1 & 0 & 0 \\
1 & 0 & -1 & 1 & 0
\end{pmatrix},
\end{equation} 
where the parameter $\alpha$ is controlling the step size of the action-based algorithm.
The ${\rm diag}({\rm\bf{v}}_{E_C})$ in Eq.(\ref{eq:act_array}) is a diagonal matrix whose entries are the approximate charging energies for plunger gates $P_1$ and $P_2$ determined during bootstrapping, ${\rm\bf{v}}_{E_C}=(V_{P_1}^{E_C}, V_{P_2}^{E_C})$. 
The definition of the action array, $\mathbb{A}$, is rooted in the topology of the double-QD device state space [see Fig.~\ref{fig:fig-1}(a)].
For any combination of global states detected by the classifier, the action vector ${\rm\bf{v}}_{act}$ is determined by taking the average action on a given gate weighted by the estimated percentage of each state in the probability vector ${\rm\bf{p}}$.
For example, when the device is assessed as being entirely in a single global state, the gate adjustment results in a simplified action assigned to that state, e.g., for ${\rm SD}_C$ [teal arrows in Fig.~\ref{fig:fig-1}(a)], ${\rm\bf{v}}_{\rm 
act}=-\alpha(V_{P_1}^{E_C}, V_{P_2}^{E_C})$.
If the predicted state indicates a transition between two global states, e.g., ${\rm\bf{p}}(x_{\rm c})=(0, 0, 0.5, 0.5, 0)$ (50~\% ${\rm SD}_C$ and 50~\% ${\rm SD}_R$), the action vector ${\rm\bf{v}}_{act}$ would be defined as the average of the per state actions: ${\rm\bf{v}}_{\rm act}=\alpha(0, -0.5V_{P_2}^{E_C})$, as shown with blue arrows in the bottom right of Fig.~\ref{fig:fig-1}(a).

The tuning process is repeated until $\delta_{tr}$ is surpassed and the algorithm declares the DD state or a tolerance error is raised.
The tolerance error indicates that the algorithm did not find a region in the plunger-plunger space that would meet the threshold requirements and results in an adjustment of the middle barrier gate.
Finally, if the algorithm tries to go out of safety limits, the tuning process is terminated, an ``out-of-bounds'' termination is recorded, and the algorithm is reinitiated. 

\subsection{Ray-based charge-tuning module}
\label{ssec:rbct}
The ray-based charge-tuning module relies on a series of 1D measurements (rays) acquired in a head-to-tail sequence and conventional data processing techniques to identify and locate charge transitions within each ray.
Navigation from an arbitrary double-QD charge state to a desired double-QD charge configuration $(m,n)$, with $m$ and $n$ indicating the number of charges accumulated on the left and right QD, respectively, is also guided by a set of physics-inspired actions.
The charge-tuning module proceeds in three steps: unloading the QD device of all charges [depicted in Fig.~\ref{fig:fig-1}(b)], navigation to the intersection of the first charge transitions and calibration of virtual gates [Fig.~\ref{fig:fig-1}(c)], and loading to a desired $(m,n)$ charge state [Fig.~\ref{fig:fig-1}(d)].
Here, we present an overview of what these steps encompass and the desired output for each of them. 

\subsubsection{Step 1: Unloading the QD}
The goal of the first step of charge tuning is to remove all charges from a double-QD.
Starting at a point $x_{\rm c}$ in the DD state where the action-based algorithm terminated, the unloading process initiates a series of ray-based measurements, each combined with a peak-finding algorithm to check for the presence of charge transitions.
Figure~\ref{fig:fig-1}(b) depicts examples of unloading paths for two sample points.
The rays' directions alternate for consecutive measurements between $-V_{P_1}$ and $-V_{P_2}$, with the initial direction chosen randomly.
The length of the rays, independent for each direction, is proportional to $V_{P_i}^{E_C}$ and is chosen to ensure that a transition is captured as long as the QD is not empty.
If at any point the ray extends past the safety limits, its length is reduced to remain within bounds.

\begin{figure}[t]
\centering
\includegraphics[width=\columnwidth]{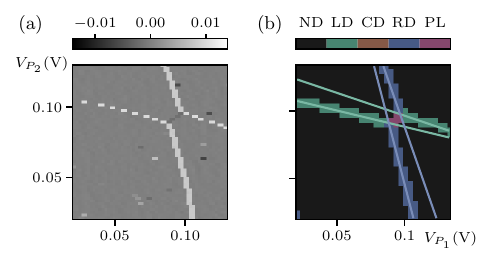}
\caption{
(a) Numerical derivative of a simulated 2D measurement, capturing the intersection of the first charge transitions, used to establish the virtual gate space.
(b) The outcome of the ML-based pixel classifier, with each pixel in the original 2D scan classified as no, left, central, or right transition or a polarization line (NT, LT, CT, RT, or PL, respectively).
}
\label{fig:fig-2} 
\end{figure}

The starting point for the consecutive measurements is determined based on the location of the last peak detected in a given ray.
The peaks are identified using either a conventional peak-finding algorithm~\cite{SciPy} or ML techniques~\footnote{Both methods are implemented in the current version of the PIT algorithm.}.
If at least one peak is detected in a ray, the consecutive measurement is initiated past the last detected peak.
If there are no peaks in a given ray,  the direction is flagged as ``potentially empty''.
In cases where the starting point for the consecutive ray falls outside of the safety limits, the direction is flagged as ``soft out-of-bounds'' and the subsequent measurement is initiated at the same point.

To prevent premature termination due to missed transitions, the algorithm terminates only when both directions are consecutively flagged as potentially empty. 
Moreover, if both directions are flagged as soft out-of-bounds in consecutive measurements, the unloading is marked as ``hard out-of-bounds.''

\subsubsection{Step 2: Establishing controllability}
In the presence of capacitive cross-talk between the various gate electrodes, changing voltages on a single gate affects not only the parameter it is designed to control (e.g., the electrochemical potential of a specific QD) but also other parameters (e.g., the electrochemical potential of the neighboring QDs and the tunnel barrier between adjacent QDs).
To enable targeted control of specific QDs and to fill a QD array into a desired charge configuration, we employ virtual gates, i.e., linear combinations of multiple gate voltages chosen to address only a single electrochemical potential~\cite{Hensgens17-FHQ, Hensgens18-PhD, Perron15-QSB}.

The emptying process, performed in the physical gate voltage space, results in points distributed within a range of distances from the first charge transition.
Given the local relevance of virtual gates and the necessity of accounting for transitions from both QDs for proper virtualization, it is desirable to navigate near the intersection of the first charge transitions prior to determining the capacitive coupling between the QD gates.
This is achieved in PIT through a simple feedback process involving a series of ray measurements and position adjustments until both transitions are located within the same distance from each ray's tail, as depicted in Fig.~\ref{fig:fig-1}(c).

Once the intersection of the final charge transitions is located, a 2D measurement capturing the intersection of transitions is performed; see Fig.~\ref{fig:fig-2}(a).
PIT leverages a ML-based pixel classifier~\cite{Lin16-FPN} and conventional linear regression to determine virtual gates~\cite{Ziegler23-AEC}.
The pixel classifier is trained on simulated data to flag every pixel in the numerical gradient of the scan as either no, left, central, or right transition (NT, LT, CT, or RT, respectively), or as a polarization line (PL). 

The resulting contiguous regions of pixels are then fit by linear regression independently for each class
This process yields images with classified pixels and corresponding fits, as shown in Fig.~\ref{fig:fig-2}.
Fitting the regions of LT and RT classes yields the off-diagonal terms of the capacitive coupling matrix, normalizing each row such that the diagonal terms are equal $1.0$~\cite{Mills19-SSC}.
If multiple transitions are present within a class, the slopes are combined by an average weighted by the standard deviations of the fits.
This method has the added benefit that the fits also give confidence intervals, and the labeled pixels give the locations of various key features of a scan.

\subsubsection{Step 3: Loading to \texorpdfstring{$(m,n)$}{} state}
The virtual gates information enables targeted control of the electrochemical potential of individual QDs.
PIT sequentially loads each QD to the desired charge state.
For example, the yellow path in Fig.~\ref{fig:fig-1}(d) shows loading one charge to each QD, with the left QD being loaded first. 
Similarly, the red path in Fig.~\ref{fig:fig-1}(d) shows loading to the $(2,1)$ state, with PIT first loading two charges to the left QD, and then one electron to the right QD.

The loading process, performed in the virtual gate space $(V_{P'_1}, V_{P'_2})$, starts at the same point where the 2D measurement used in step 2 was initiated.
At each step, a ray of length approximately $3V_{P_i}^{E_C}$ is measured, followed by the peak-finding module.
If a peak is not found, the starting point for the next ray, $x_{\rm c}$, is set at $20~\%$ before the end of the current ray and the measurement is repeated~\footnote{The $20~\%$ of ray length requirement is determined by the lower bound on the distance from the peak necessary to properly identify it.}.
The consecutive measurements are initiated at the midpoint between the first two peaks in a given ray.
If only one peak is found, the measurement is repeated.
Once the first QD is loaded with the required number of charges, the loading direction is switched. 
To prevent a ray from going through a polarization instead of a transition line, a pair of auxiliary rays orthogonal to the loading direction is measured once at least one electron is loaded on each QD.
These are used to reposition the starting points of the consecutive rays within the honeycomb; see Appendix~\ref{app:ray-adj} for details.

Once PIT declares that the desired charge configuration has been established, a final check involving a pair of long rays measured in the unloading directions $(-V_{P'_1},-V_{P'_2})$ is executed. 
The length of rays used in the check is set to $(m+1)V_{P_1}^{E_C}$ and $(n+1)V_{P_2}^{E_C}$ for the $V_{P'_1}$ and $V_{P'_2}$, respectively, and $(n,m)$ the target charge configuration.
If the expected number of peaks is detected in each ray, PIT terminates. 
Otherwise, the full ray-based charge-tuning module is reinitiated from the current point, up to a maximum of three times.

\subsection{Data}
\label{ssec:data}
Both modules of the PIT algorithm are developed using a set of ten simulated double-QD devices with a similar gate architecture and varying levels of noise~\cite{Ziegler22-TRA}. 
An ensemble of additional seven qualitatively distinct simulated double-QD devices with varying levels of noise is used to test the performance of PIT.
The benchmark noise level used in the simulation is established by applying a quantitative version of a DQC module~\cite{Ziegler22-TRA} to a dataset of 756 small experimentally measured 2D scans available via the National Institute of Standards and Technology (NIST) Science Data Portal~\cite{qf-data}.

The reliability of PIT is further validated using a set of 23 large 2D experimentally measured scans: 7 included in the {\it QFlow 2.0: Quantum dot data for machine learning}~\cite{qf-data} dataset and 16 new scans acquired using two double-QD configurations on two different three-QD Si$_x$/SiGe$_{1-x}$ devices, fabricated on an industrial 300-\si{\milli\meter} process line~\cite{Pillarisetty21-SQC}.
Electrostatic variations to the exact double-QD configuration were induced by changing the middle barrier and other adjacent gates for each scan.  

The simulated double-QD devices are also used to test the design of an autonomous tuning process that incorporates both PIT and DQC~\cite{Ziegler22-TRA}.
In those tests, the noise level is adjusted throughout the tuning process in a feedback loop in response to the DQC to mimic an online experimental test.

\section{Results}
\label{sec:results}
To validate the PIT algorithm, we use a set of qualitatively distinct simulated double-QD devices with varying levels of noise as well as two sets of large experimentally acquired 2D scans discussed in Sec.~\ref{ssec:data}. 
The noise levels used in all tests with simulated data are varied around a reference level of noise extracted from the experimental data that, for simplicity, we denote as $1.00$~\footnote{The noise distribution extracted from 756 small experimental scans ranges from $0.05$ to $5.35$ in the units of the optimized noise configuration from Ref.~\cite{Ziegler22-TRA}, with the noise level mean $\mu=0.3(4)$ and median $M=0.2$ (median absolute deviation MAD=0.07). 
Since the data is highly skewed, with the Fisher-Pearson coefficient of skewness $g_1=9.1$, we opt to use the median as a measure of central tendency for the noise distribution.}.

In all tests, the tuning runs are executed automatically in a sequential manner, with the consecutive modules and steps being initiated only for points that tuned successfully at the preceding step.
The success rates reported throughout this section are determined based on the number of points for which each step was initiated to facilitate an unbiased comparison of the performance of PIT's individual components.
Given the high success rate for both tuning to the DD state and emptying steps, including all points would not significantly change the results.

\subsection{Algorithm configuration}
\label{ssec:config}
The configuration of PIT used in this work is established based on a combination of prior work and tests on an ensemble of ten qualitatively similar double-QD simulations with varying noise levels~\cite{Ziegler22-TRA}.
The termination of the action-based navigation algorithm in the coarse-tuning module relies on a distance function that approximates the distance from a measurement point to the target double QD, with a penalty function to further encourage termination away from undesired states, as proposed in Ref.~\cite{Zwolak20-AQD}.
The choice of the exit threshold $\delta_{\rm tr}$ is determined based on the analysis of the simulated data and translates directly to how close in the state space the final point is to the DD region: $\delta_{\rm tr}=0.3$ indicates at least $95~\%$ of the DD state in the captured measurement while for $\delta_{\rm tr}=0.9$ the algorithm will terminate as soon as the DD prediction surpasses $55~\%$.
The $\delta_{\rm tr}$ level can be made more or less restrictive, depending on the expected coupling of the double-QD system.
PIT assumes by default $\delta_{\rm tr}=0.4$ (which translates to about $85~\%$ of the DD requirement).

The action-based tuning requires a single parameter defining the scale for the actions.
For this, we set the scale $\alpha = 1.5V^{E_C}$ so that each step does not move into a completely unknown area.
For the ray-based charge tuning, there are similar parameters defining the relative lengths of rays at each stage.
By default, the length of rays is $\ell_{\rm empty} = 1.5V_{P_i}^{E_C}$ for the emptying step, $\ell_{\rm control} = 2.25V_{P_i}^{E_C}$ for the controllability step, and $\ell_{\rm load} = 3V_{P_i}^{E_C}$ for the loading step, with $P_i$, $i=1,2$ indicating the plunger gates.
These lengths are chosen such that the number of data points measured is balanced with the importance at each stage of finding a peak should it exist.
On unloading it is assumed that a ray with no peaks indicates the direction is empty.
Loading rays are required to have one peak in order to make a step.

In all tests we are using the conventional peak-finding algorithm~\cite{SciPy}.
Thus, the last major parameter for ray-based charge tuning is the prominence of expected peaks.
This prominence must be set to account for variations in peak height that might result from imperfect sensor compensation or increased tunneling at high charge occupations.
Here, we set a prominence equal to $1/3$ of the maximum estimated from all rays measured in a given tuning run.
This maximum is initiated using the final measurement of the action-based tuning stage and updated using a Metropolis-Hastings algorithm~\cite{Hasting70-MCS}.

\subsection{Benchmarking with simulated data}
\label{ssec:perf-sim}
Testing PIT on simulated data facilitates a controlled study of how the various types, combinations, and prevalence of noise impact the functioning of the tuner. 
The spectrum of device designs used in simulations is chosen to capture the effective realization of edge cases, such as strongly and weakly coupled QDs. 
In our preliminary tests we found that varying the ``optimized'' combination of noise established in Ref.~\cite{Ziegler22-TRA} results in a much higher likelihood of the telegraph noise than would be expected in typical experiments.
This is likely because the noise optimization in Ref.~\cite{Ziegler22-TRA} was performed using moderate to low-quality experimental data to ensure that the state classifier is reliable and robust.
Thus, in all tests with simulated data, the lifetimes and magnitudes for the telegraph noise are kept fixed at a level consistent with experimental data determined through the ML assessment of experimental devices and visual inspection to be a $2~\si{\second}$ lifetime for both the upper and lower states (assuming a $10~\si{\milli\second}$ integration time) with a magnitude set to 4 times the relative magnitude used in Ref.~\cite{Ziegler22-TRA} and the magnitudes for the telegraph, $1/f$ (pink), and white noise are varied between $0.2$ and $2.5$ of the reference noise level.
Additional results for varying levels of the telegraph noise are included in Appendix~\ref{app:per-sm}. 

The three panels in Fig.~\ref{fig:tuning-sim} show the performance of the individual PIT components.
The action-based coarse tuning to the DD state module seems to be quite robust against noise, with an overall performance of $94.6(2.9)~\%$ when using ray-based measurements and $98.9(2.1)~\%$ when tuning with small 2D scans~\footnote{We use a notation value(uncertainty) to express uncertainties, for example, $1.5(6)~\si{\centi\meter}$ would be interpreted as $(1.5\pm0.6)~\si{\centi\meter}$. 
All uncertainties herein reflect the uncorrelated combination of single-standard deviation statistical and systematic uncertainties.}.

An analysis of failed tuning paths suggests that the slightly worse performance of the ray-based tuning is likely caused by the design of the action-based tuning that relies on the assumption that scans capturing transition between states are properly quantified by partial labels returned by the classifier.
In other words, there is a smooth change in the predicted probability with incremental gate adjustments. 
The RBC, on the other hand, predicts the most likely state for a given point, which results in more abrupt changes in the probability vector.
This makes the action-based navigation between states a little more challenging.  

\begin{figure}[t]
\centering
\includegraphics[width=\linewidth]{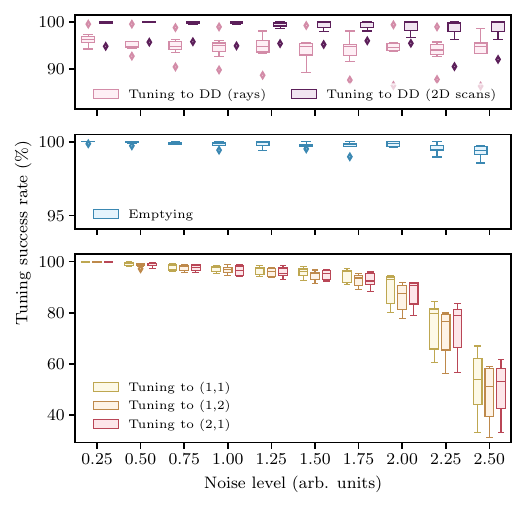}
\caption{
Performance on simulated data for varying noise levels, with each panel representing a consecutive step PIT takes.
Both the action-based coarse tuning (top panel) and emptying (middle panel) show consistently high success rates regardless of the noise level.
The performance of the charge-setting module deteriorates slowly as the noise level increases, with a significant drop at around $2.0$ (i.e., double the noise level estimated from experimental data).
}
\label{fig:tuning-sim} 
\end{figure}

\begin{figure*}[t]
\centering
\includegraphics[width=\linewidth]{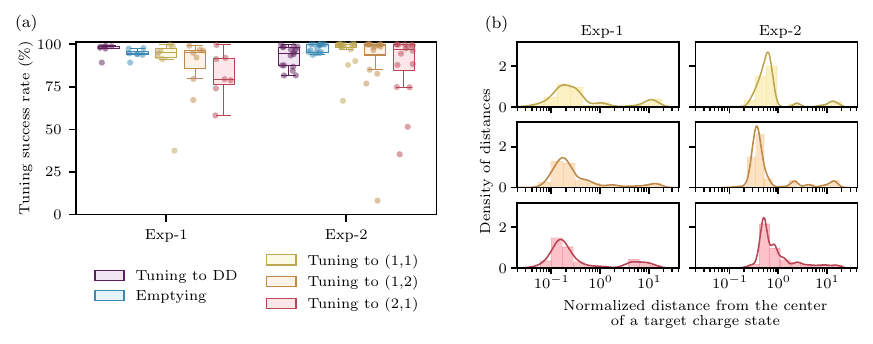}
\caption{
Performance of PIT algorithm on experimental data.
(a) A box plot showing the off-line performance of the individual components with three target charge-state configurations.
The central lines indicate medians for each test and the central box represents $50~\%$ of the data.
The whiskers extend to either the extreme values or $1.5$ times the interquartile range, whichever is closer to the median.
The individual points on top of each box plot show the success rates for each device.
(b) Histogram of normalized distances from the center of the target honeycombs, with a kernel density estimate curve overlaid.
The distances are normalized on a per device basis by the radius of an inscribed circle centered within the target charge state.
Distances of no more than one guarantee success.  
These distributions confirm that PIT not only reliably terminates in the desired charge configuration, but also converges well within the target charge hexagon.
}
\label{fig:tuning-exp} 
\end{figure*}

For comparison, coarse tuning to the DD using Nelder-Mead optimization~\cite{Nelder65-NMA,Gao12-IMN,Zwolak20-AQD} results in significantly lower success rates of $58.5(7.6)~\%$ with rays and $82.5(6.3)~\%$ with 2D scans.
The number of measurement iterations for the action-based tuning is reduced by a factor of 2 for ray-based coarse tuning [$5.3(1.2)$ versus $13.3(1.8)$ with Nelder-Mead optimization] and by a factor of 3 for tuning with 2D scans [$4.2(6)$ versus $12.9(2.0)$ with Nelder-Mead optimization].
This corresponds to an average overall data reduction at the DD tuning stage of about $60~\%$ for action-based coarse tuning with rays and around $67~\%$ data reduction for tuning with 2D scans.

The performance for the emptying step, shown in the middle panel of Fig.~\ref{fig:tuning-sim}, is also quite high, with an overall $99.8(3)~\%$ success rate.
However, we observe that the likelihood of the soft out-of-bounds termination increases with noise, from $2.6(3.6)~\%$ at a $0.25$ noise level to $21.5(10.2)~\%$ at noise level $2.50$.  
The likelihood for hard out-of-bounds termination is below $1~\%$ regardless of the noise level.

For setting a desired charge configuration, depicted in the bottom panel of Fig.~\ref{fig:tuning-sim}, there is a correlation between the performance of PIT and the noise level, as expected.
The success rate decreases steadily with the increasing noise up to the noise level of $1.75$ and then drops rapidly once the noise level surpasses about $2.00$.

\subsection{Off-line tuning with PIT}
\label{ssec:perf-exp}
Tuning off-line---that is tuning within large experimentally measured 2D scans capturing multiple state configurations---enables validating PIT in the presence of real-world noise and implications. 
PIT is tested on two experimentally measured sets of scans.
The first set, \emph{exp-1}, consists of seven scans from the {\it QFlow 2.0: Quantum dot data for machine learning}~\cite{qf-data}. 
These scans are measured over a fixed voltage range for the plunger gates ($150~\si{\milli\volt}$ to $550~\si{\milli\volt}$ for the first gate and $100~\si{\milli\volt}$ to $500~\si{\milli\volt}$ for the second gate). 
The second set, \emph{exp-2}, includes 16 scans, ranging in size from $400~\si{\milli\volt}$ by $400~\si{\milli\volt}$ to $600~\si{\milli\volt}$ by $600~\si{\milli\volt}$.
The performance of PIT on those two datasets is shown in Fig.~\ref{fig:tuning-exp}(a).
Since the measurement design for RBC implemented in PIT is not compatible with the static off-line scans, coarse-tuning tests with experimental data are done only using the 2D scans. 

The starting points for tests within experimental scans are sampled in a grid with an exclusion of regions where the signal-to-noise ratio (SNR) is insufficient.
The exclusion regions are determined prior to testing based on visual inspection of the data combined with analysis of the distribution of the charge sensor response for a series of small 2D scans densely sampled within the large scan~\cite{Zwolak21-RBI}.
In an online implementation, the data quality control module~\cite{Ziegler22-TRA} would initiate a charge sensor recalibration. 
In an off-line setting, such corrective actions are not possible (see Sec.~\ref{ssec:tuning-playground} for details).
The overall number of points sampled is on average $1\,800(16)$ for the exp-1 set while, for the exp-2 set, it varies between $346$ and $6\,124$.
The larger variability in the number of points per scan for the exp-2 set is due to varying scan voltage ranges, sampling initiation points at constant density in each scan, and excluding points from poorly charge-sensed regions.
The bounds for off-line tuning are offset with respect to safety limits to ensure that the initial measurements can be sampled.
However, during tuning, the algorithm can navigate toward the safety limits.
If at any point the algorithm suggests going out of safety bounds, the tuning process is terminated and an out-of-bounds termination failure is recorded. 
The success rate for the action-based coarse tuning to the DD state using small 2D scans is slightly higher for exp-1, at $97.1(3.5)~\%$, than for exp-2, at $92.5(6.5)~\%$.
For comparison, the success rate for tuning to a double-QD state reported in Ref.~\cite{Darulova19-ATQ} was $80~\%$ for a set of five double-QD devices over two thermal runs.
Tuning using the Nelder-Mead optimization~\cite{Zwolak20-AQD} results in success rates of $76.8(4.0)~\%$ for exp-1 and $83.5(15.7)~\%$ for exp-2.
Similar to benchmarking with simulated data, the number of iterations is also about 2 to 3 times higher for Nelder-Mead optimization than for the action-based tuning, with $11.7(6)$ versus $3.9(2)$ for exp-1 and $8.4(4.0)$ versus $3.9(1.4)$ for exp-2.

A post-testing analysis of the failing cases reveals that one of the two main causes for the action-based tuning failure is a repeated incorrect assessment of the state vector ${\rm\bf{p}}(x_{\rm c})$ by the ML module, which is especially prevalent in scans containing poorly-sensed regions.
The worst case of this failure mode observed in the off-line tests is a true DD state identified as RD, which results in an action that is the opposite of the true best action.

The second failure mode, relevant only to off-line testing, is related to the out-of-bounds termination: tuning runs are declared as failures whenever a measurement would surpass the scan limits.
This is particularly problematic for scans where the DD region reaches the edge of the scan.
Out of the 16 large scans in exp-2, six fall into either one or both of these categories and have an overall success rate significantly lower than for the remaining scans, at $85.0(2.8)~\%$ versus $97.1(2.2)~\%$.

There are a number of features implemented in PIT that are likely to significantly lower the likelihood of the out-of-bound termination but that could not be tested off-line.
For example, an incorrect state assessment can be minimized with charge sensor recalibration based on quality control outputs~\cite{Ziegler22-TRA}.
To handle issues beyond quality, adjustment of the exit threshold $\delta_{tr}$ for cases with strongly coupled QDs and a more sophisticated logic incorporating the memory of state measurements would likely improve the action-based tuning performance.
However, an analysis of the out-of-bounds termination cases for the off-line tests allows for a clearer understanding of the possible failure modes.
This, in turn, can inform the design of the bootstrapping module to ensure the initialization of PIT in a configuration that gives the highest probability of success in an online deployment.

For the emptying step, the $(0,0)$ state can be reached in three ways, with a proper and soft out-of-bounds termination (both considered a tuning success), and a hard out-of-bounds termination (considered a tuning failure).
For this step, the success rate for exp-1 is $94.5(2.8)~\%$ [with $10.4(10.5)~\%$ rate of soft out-of-bounds] and $97.9(2.5)~\%$ for exp-2 [with $6.8(9.9)~\%$ rate of the soft out-of-bounds].
The hard out-of-bounds failure rate is $0.1(2)~\%$ and $0.1(1.2)~\%$ for exp-1 and exp-2, respectively.
For comparison, the success rate for ML-driven emptying reported in Ref.~\cite{Durrer19-ATQ} was at $90~\%$ over 160 online experimental runs.

\renewcommand{\arraystretch}{1.2}
\renewcommand{\tabcolsep}{2pt}
\begin{table}[b]
    \caption{Summary of tuning success statistics for the two experimental datasets, with the standard deviation (st. dev.) and median absolute deviation (MAD) given in parentheses.
    }
    \label{tab:exp-charge-tuning}
    \centering
    \begin{ruledtabular}
    \begin{tabular}{llccc}
        && \multicolumn{3}{c}{Target charge state} \\
        Set & Statistics & $(1,1)$ & $(1,2)$  & $(2,1)$ \\ \hline
        Exp-1& Mean (st. dev.) & 87.3(22.2) & 89.5(11.7) & 81.9(13.8) \\
        & Median (MAD) & 95.2(14.2) & 94.9(9.2) & 79.5(10.6) \\ [1.2ex]
        Exp-2 & Mean (st. dev.) & 94.9(10.9) & 88.8(23.3) & 87.1(18.9) \\
        & Median (MAD) & 99.6(7.2) & 98.9(13.9) & 96.3(13.1)  \\
    \end{tabular}
    \end{ruledtabular}
\end{table}

The summary statistics for the final stage of PIT, i.e., setting a target charge configuration, are presented in Table~\ref{tab:exp-charge-tuning}. 
Given the long tails of the charge-tuning success rate distributions as well as the presence of outliers, in addition to the overall average performance per target charge configuration, we report also the central tendency and dispersion of performance using median and median absolute deviation.
We find that the success rate is fairly comparable between the two datasets, with tuning to the $(2,1)$ charge state being less successful than tuning to $(1,1)$ and $(1,2)$.
This asymmetry in charge-tuning performance is likely due to the direction in which the data is acquired, with the fast scan direction typically giving a cleaner derivative than the slow scan direction, combined with the effect of off-line virtualization. 

Distributions of the final positions relative to the center of the target charge state across each experimental dataset can be found in Fig.~\ref{fig:tuning-exp}(b).
In order to compare performance between devices and charge states we normalize these distances by the radius of an inscribed circle for the target charge region.
In these units a distance of no more than $1.0$ guarantees success.
An analysis of these distributions confirms that the final positions end up close to the center of the target region.
Failures fall into two roughly equal-sized groups: those roughly one charge away and those far away.
The nearby failures are primarily due to a single transition being missed on loading.
The farther failures tend to be due to poor SNR causing excessive peaks to be found [ending in the $(0,0)$ state] or very few peaks to be found (ending at large charge occupation).
The failure cases are especially apparent in the success rates of the outliers in Fig.~\ref{fig:tuning-exp}, and are compounded by off-line scans exhibiting static noise that can potentially affect many tuning runs.

In order to overcome some of these limitations we may apply a number of improvements.
Incorporating a quality assessment of rays to detect poor SNR could reduce the failures we see here.
We may also use methods to reduce the impact of low SNR by taking ray measurements multiple times and comparing peaks found to ensure more robust peak identification.
For comparison, the success rate for a ML-driven online charge setting with the target state chosen randomly from a set of four possible configurations reported in Ref.~\cite{Durrer19-ATQ} was about $63~\%$.

\subsection{Simulated tuning with noise adjustment}
\label{ssec:tuning-playground}
To better estimate the expected \emph{in situ} performance of PIT, we implement a tuning framework proposed in Ref.~\cite{Ziegler22-TRA} using the QD simulator~\cite{Zwolak18-QLD}.
To ensure that only reliable data is analyzed by the ML state classifier, we incorporate the DQC system into the action-based coarse-tuning module and add a recalibration step that is executed whenever the data is assessed as either moderate or low quality.
Whenever recalibration is initiated, the device noise level is adjusted and another measurement is taken at the same point.
In our simulated tests, we set each call of the recalibration function to reduce the noise level by $30~\%$ of its current value.
The process is repeated until the scan is deemed suitable for further processing.
The purpose of the DQC system is to prevent the use for tuning data that would result in poor state estimation and, ultimately, a failure of the charge-tuning module.
In an online implementation, it may be preferable to terminate tuning of the device altogether for a low-quality data assessment or if repeated sensor recalibration does not improve data quality.

\begin{figure}[t]
\centering
\includegraphics[width=\linewidth]{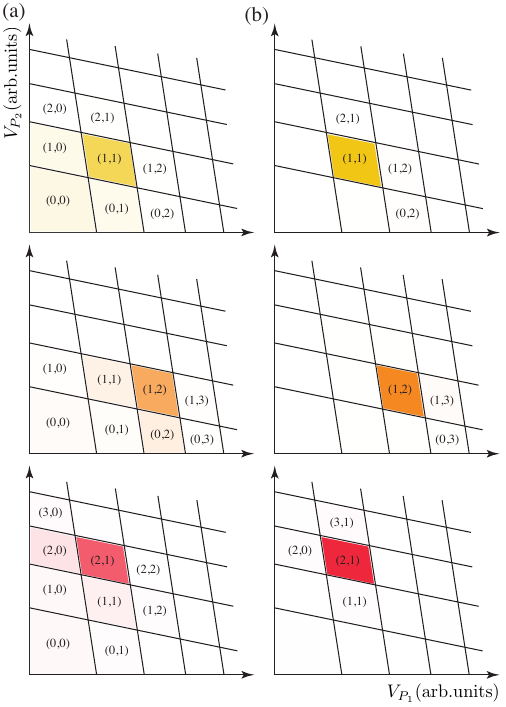}
\caption{
The results of the simulated tuning with noise adjustment test runs (a) using a single ray per step during the loading process and (b) following the ``repeated measurement with voting'' strategy with three repeats, with the latter showing a clear improvement in performance.
The target states are $(1,1)$, $(1,2)$, and $(2,1)$ for the top, middle, and bottom rows, respectively.
The palette corresponds to colors used in Fig.~\ref{fig:tuning-sim} and the intensity indicates the frequency of terminating at a particular configuration, averaged over all devices. 
In all cases, we label only those states at which a given series of test runs terminated.
}
\label{fig:tuning-playground} 
\end{figure}

In all tests we assume that the SNR varies linearly with respect to each plunger gate, setting the maximum noise level for each scan to $30$ times the reference noise, the minimum to $1.00$, and an equal slope with respect to both gates to mimic an imperfectly compensated charge sensor.
This range for noise magnitudes sets the center of each device to have a noise level roughly equivalent to the moderate quality in Ref.~\cite{Ziegler22-TRA}.
Using the set of seven simulated devices, we initiate 100 tuning runs per device, with the starting points randomly sampled within each scan.
To verify the utility of the DQC and recalibration components of the tuning framework, we test tuning on devices with the same noise characteristics both with and without quality-aware noise adjustment.
We find that even without noise adjustment the action-based tuning module is quite successful at $88.9(8.5)~\%$.
However, setting any target charge configuration fails almost completely with an overall $0.1(6)~\%$ success rate.
With noise adjustment, the success of the action-based tuning remains high at $88.3(4.6)~\%$, but the tuning success rate increases for $(1,1)$ to $70.2(24.4)~\%$, for $(1,2)$ to $68.1(10.0)~\%$, and for $(2,1)$ to $71.0(14.6)~\%$.
Figure~\ref{fig:tuning-playground}(a) shows the final charge configuration distribution overlaid on a schematic stability diagram.
The color intensity corresponds to the frequency of a particular configuration averaged over all devices.
In all cases, we label only those states at which a given series of test runs terminated.

In all tests, PIT terminates at most two transitions away from the target state.
The main factor affecting the charge-tuning success rates is either missing a transition or identifying noise as transitions when loading charges.
One way to overcome this limitation is to develop a DQC module for the ray-based measurements analogous to the one used for the 2D scans during coarse tuning.
Another way to boost the performance is to implement a ``repeated measurement with voting'' strategy.
The latter approach works by determining the presence of transitions based on a repeated measurement of a ray from the same point followed by a comparative analysis of peak locations found in all rays using an algorithm that resembles a majority vote approach.

Similar to tests with a single ray, we test the repeat-and-vote strategy with three rays both without and with the noise-driven device recalibration.
Without device recalibration, the voting strategy increases the performance of tuning for $(1,1)$ to $37.1(29.8)~\%$, for $(1,2)$ to $26.0(22.1)~\%$, and for $(2,1)$ to $28.2(25.7)~\%$.
Including both data quality adjustment and three-ray voting yields charge-tuning performances of $98.0(2.9)~\%$ for $(1,1)$, $94.2(5.6)~\%$ for $(1,2)$, and $94.4(7.3)~\%$ for $(2,1)$; see Fig.~\ref{fig:tuning-playground}(b).
These results confirm that the repeated measurement with a voting strategy significantly improves the overall robustness of charge tuning against random noise.

\section{Summary and outlook}
\label{sec:conclusion}
Our results show that the PIT algorithm is very effective at device tuning with efficient use of measurements.
We show that action-based coarse tuning can navigate directly to a target global state using established ML approaches.
Moreover, this state navigation is agnostic to details of the underlying ML tool, which allows further reductions in measurement burden by using a set of 1D measurements for state estimation~\cite{Zwolak20-RBC, Zwolak21-RBI}.
The ray-based charge-tuning module enables similarly data-efficient navigation from the coarse-tuning position to a target charge state with a high rate of success.
Combining these modules gives a reliable and data-efficient algorithm for taking a device from a basic voltage estimate to a region suitable for fine-tuning and qubit operation.
Continued improvements in device quality and understanding will lead to higher success in bootstrapping methods, further improving starting points for PIT and, therefore, increasing its efficiency.

Given the success rate for our methods in off-line experimental tests as well as successes when applied to simulated data with noise adjustment, we expect that PIT will be highly effective in tuning experimental QD devices to various charge states \textit{in situ}.
An important difference between the {\it in situ} and the off-line tests is that the experimental devices tuned online are not static and can have dynamic defects that alter SNR as tuning is performed.
While this could work to our disadvantage, our demonstration of interactive tuning tests in Sec.~\ref{ssec:tuning-playground} shows how the recently developed ML tools for data quality assessment~\cite{Ziegler22-TRA} can be leveraged to flag data potentially unsuitable for ML or conventional analysis before it causes tuning failure.
For the ray-based charge tuning in an online setting, the repeat-and-vote strategy of peak finding, also described in Sec.~\ref{ssec:tuning-playground}, can be used to alleviate the effects of device noise.
While not possible during off-line testing, these simple adjustments to improve data quality and mitigate random noise, already implemented in PIT, are likely to make the {\it in situ} performance of PIT better than in off-line tests.

Although here we only consider tuning of double-QD arrays, our methods are easily generalizable to larger arrays if an $(n+1)$ loading strategy is used~\cite{Volk19-LQR}.
In such a framework, loading of an arbitrary size QD array is reduced to the tuning of only two types of 2D voltage spaces: plunger-plunger space (addressed here), and plunger-barrier space.
Extending our methods to virtual plunger-barrier space may be a relatively simple application of ray-based charge-tuning methods if reasonable estimates for barrier voltages can be made.
If more precise tunnel coupling calibration is required, additional functionality may be needed to extract transition line widths or visibility.
Such a tool may be desirable in general to incorporate a method for checking and adjusting tunnel coupling to avoid potential failure modes of our autotuning algorithm.
Future work could include using maximum entropy methods to model the probabilistic electron occupancy state from the plunger-plunger space for fine-tuning the double-QD state~\cite{Caticha22-Ent}. 
Using the tools of information geometry to measure the curvature of current transition lines in the plunger-plunger space might allow for further tunnel coupling calibration~\cite{Caticha21-Gibbs}.
The modular nature of PIT lends itself well to both additions of modules targeting tuning steps currently not incorporated in PIT (i.e., bootstrapping and fine-tuning) and further improvement to existing subroutines. 

PIT combines modern computer vision, machine learning, and data processing techniques with human heuristics to provide an intuitive, efficient, and reliable tool for QD device calibration.
Moreover, the significantly reduced one-dimensional data acquisition requirements combined with simplified data analysis techniques make PIT well suited for implementation with dedicated hardware closely integrated with the QD chip.
It is thus a major step toward fully automated and scalable tuning of QD devices, a prerequisite to using QD-based quantum computers.

\begin{acknowledgments}
This research was performed while J.Z. held an NRC Research Associateship award at NIST.
The views and conclusions contained in this paper are those of the authors and should not be interpreted as representing the official policies, either expressed or implied, of the U.S. Government. 
The U.S. Government is authorized to reproduce and distribute reprints for Government purposes notwithstanding any copyright noted herein. 
Any mention of commercial products is for information only; it does not imply recommendation or endorsement by NIST.
\end{acknowledgments}

\appendix

\section{Additional performance analysis}
\label{app:per-sm}
\begin{figure*}[t]
\centering
\includegraphics[width=\linewidth]{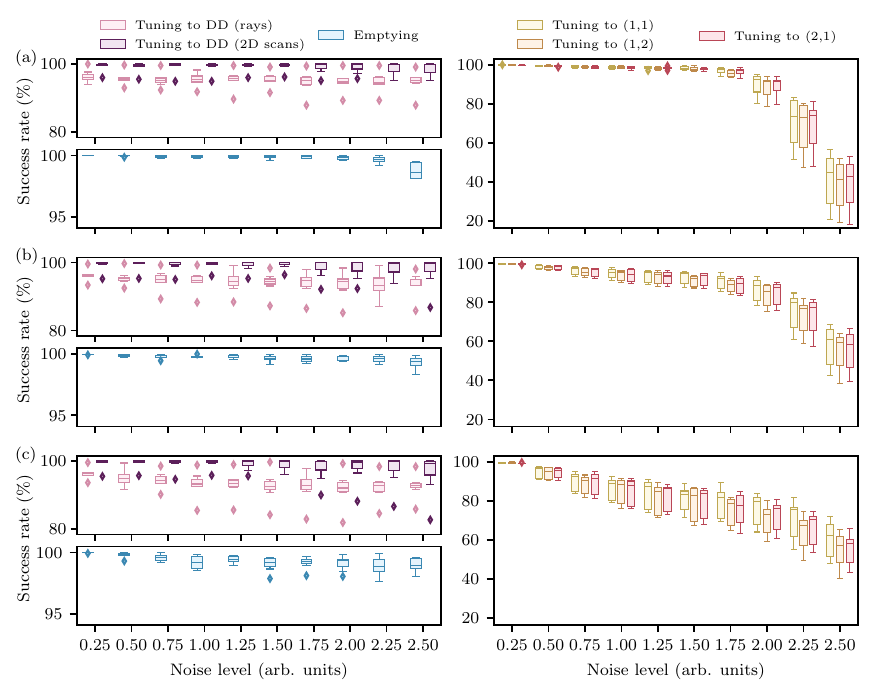}
\caption{
Additional performance tests on simulated data. 
In each panel the magnitudes for the telegraph, $1/f$ (pink), and white noise are varied between $0.2$ and $2.5$ of the reference noise level while the lifetimes for the telegraph noise are kept fixed in each panel at (a) $1~\si{\second}$ for upper and lower states (b) $2~\si{\second}$ for upper and $1~\si{\second}$ for lower, and (c) $4~\si{\second}$ lifetime for upper and $2~\si{\second}$ for the lower state (assuming an integration time of $10~\si{\milli\second}$ per pixel).
The telegraph noise magnitude is set to 4 times the relative magnitude used in Ref.~\cite{Ziegler22-TRA}.
}
\label{fig:sim-perf-comp} 
\end{figure*}

To determine the effect of telegraph noise on the tuning process, we vary the lifetime of the synthetic telegraph noise with respect to the reference noise level used in Fig.~\ref{fig:tuning-sim} and run PIT in the same manner as described in Sec.~\ref{sec:methods}.
The lifetimes of the telegraph noise used in the initial tests, summarized in Fig.~\ref{fig:tuning-sim}, were set to $2~\si{\second}$ (assuming a $10~\si{\milli\second}$ integration time) for both the upper and lower states.
For the additional tests, we consider both shorter and longer lifetimes (resulting in higher and lower amounts of telegraph noise, respectively).
The summary of the performance statistics for this data is shown in Table~\ref{tab:sim-perf-comp} and depicted graphically in Fig.~\ref{fig:sim-perf-comp}.
Figure~\ref{fig:sim-perf-comp}(a) uses a lifetime of $1~\si{\second}$ for the upper state and $1~\si{\second}$ for the lower, Fig.~\ref{fig:sim-perf-comp}(b) uses $2~\si{\second}$ for the upper and $1~\si{\second}$ for the lower state, and Fig.~\ref{fig:sim-perf-comp}(c) uses a lifetime of $4~\si{\second}$ for the upper state and $2~\si{\second}$ for the lower.

\begin{table}[b]
    \caption{Summary of tuning success statistics.
    The columns represent noise combinations depicted in Fig.~\ref{fig:sim-perf-comp}.
    All results are reported as mean(st.dev.).
    }
    \label{tab:sim-perf-comp}
    \centering
    \begin{ruledtabular}
    \begin{tabular}{lrrr}
        Tuning phase & (a) & (b) & (c)\\ \hline
        DD (rays) & 95.3(2.6) & 94.3(3.2) & 93.3(3.9) \\
        DD (scans) & 99.0(1.7) & 98.7(2.5) & 98.4(3.4) \\
        Emptying & 99.8(4) & 99.7(3) & 99.3(6) \\
        Soft out-of-bounds & 8.3(9.4) & 11.7(9.3) & 18.6(12.0) \\
        Hard out-of-bounds & 0.2(4) & 0.2(3) & 0.5(5) \\
        Tuning to $(1,1)$  & 89.0(19.4) & 88.3(13.4) & 81.7(13.0) \\
        Tuning to $(1,2)$ & 88.2(20.1) & 86.8(14.1) & 78.8(14.6) \\
        Tuning to $(2,1)$ & 88.4(19.9) & 87.1(14.1) & 79.4(14.3) \\
    \end{tabular}
    \end{ruledtabular}
\end{table}

Regardless of the telegraph noise lifetime, the performance of action-based tuning is remarkably high, at about $94~\%$ when using rays and about $99~\%$ when using 2D scans; see top two rows in Table~\ref{tab:sim-perf-comp}.
The emptying success rate is also consistently over $99~\%$, though the soft out-of-bounds rate increases from an average of $8.3(9.4)~\%$ for short lifetimes of telegraph noise to $18.6(12.0)~\%$ for long lifetimes of telegraph noise.
The hard out-of-bounds remains consistently below $1~\%$.
This is an important feature since the hard out-of-bounds termination indicates a complete failure to recognize an empty state.

For charge tuning, there is a somewhat unexpected trend of rapidly decreasing performance at high noise levels for the short lifetimes, Fig.~\ref{fig:sim-perf-comp}(a), that can be explained by the algorithm's inability to ignore a telegraph jump in a ray when a chance of such jumps is unusually low.
This issue is likely addressable by using the repeat-and-vote strategy described in Sec.~\ref{ssec:tuning-playground}, or by making the peak-finding algorithm more restrictive by increasing the expected prominence of peaks.
For moderate but higher than the reference frequency of telegraph noise, Fig.~\ref{fig:sim-perf-comp}(b), the performance of PIT is nearly indistinguishable from that shown in Fig.~\ref{fig:tuning-sim}.
Finally, the performance degrades substantially when the lifetime of telegraph noise becomes significantly longer than the reference. 

\begin{figure*}[t]
\centering
\includegraphics[width=\linewidth]{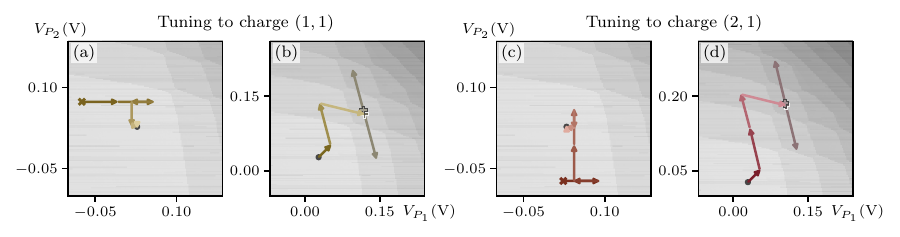}
\caption{
Detailed paths of the controllability and loading process depicted in Figs.~\ref{fig:fig-1}(c) and \ref{fig:fig-1}(d) showing the auxiliary rays.
(a),(c) Two paths for navigating to the point where virtual gates may be established, with an ``x'' indicating the beginning of the path and the termination point marked with a black dot.
(b),(d) Two paths navigating to the target charge state. 
The desaturated arrows indicate the ray measurements used for the location adjustment, with the white-bordered ``+'' indicating the location before adjustment and the black-bordered ``+'' showing the position after.
}
\label{fig:final-loc-adjust} 
\end{figure*}

\section{Location adjustment}
\label{app:ray-adj}
Figure~\ref{fig:final-loc-adjust} shows more detailed paths for the controllability and loading steps of PIT for setting charge $(1,1)$ and $(2,1)$ as those shown in Fig.~\ref{fig:fig-1}(c) and (d), depicted on separate plots for better clarity.
The dark, desaturated arrows shown in Fig.~\ref{fig:final-loc-adjust}(b) and (d) indicate the auxiliary rays used to make the location adjustment.
The positions before and after this adjustment are indicated by the light and dark ``+'', respectively.
This adjustment helps to ensure that the consecutive measurement is made in the center of a hexagon edge and that the termination point is well centered in a charge region.
The orientation of the arrows here corresponds to the virtual gate space.

%


\begin{thebibliography}{45}%
\makeatletter
\providecommand \@ifxundefined [1]{%
 \@ifx{#1\undefined}
}%
\providecommand \@ifnum [1]{%
 \ifnum #1\expandafter \@firstoftwo
 \else \expandafter \@secondoftwo
 \fi
}%
\providecommand \@ifx [1]{%
 \ifx #1\expandafter \@firstoftwo
 \else \expandafter \@secondoftwo
 \fi
}%
\providecommand \natexlab [1]{#1}%
\providecommand \enquote  [1]{``#1''}%
\providecommand \bibnamefont  [1]{#1}%
\providecommand \bibfnamefont [1]{#1}%
\providecommand \citenamefont [1]{#1}%
\providecommand \href@noop [0]{\@secondoftwo}%
\providecommand \href [0]{\begingroup \@sanitize@url \@href}%
\providecommand \@href[1]{\@@startlink{#1}\@@href}%
\providecommand \@@href[1]{\endgroup#1\@@endlink}%
\providecommand \@sanitize@url [0]{\catcode `\\12\catcode `\$12\catcode
  `\&12\catcode `\#12\catcode `\^12\catcode `\_12\catcode `\%12\relax}%
\providecommand \@@startlink[1]{}%
\providecommand \@@endlink[0]{}%
\providecommand \url  [0]{\begingroup\@sanitize@url \@url }%
\providecommand \@url [1]{\endgroup\@href {#1}{\urlprefix }}%
\providecommand \urlprefix  [0]{URL }%
\providecommand \Eprint [0]{\href }%
\providecommand \doibase [0]{https://doi.org/}%
\providecommand \selectlanguage [0]{\@gobble}%
\providecommand \bibinfo  [0]{\@secondoftwo}%
\providecommand \bibfield  [0]{\@secondoftwo}%
\providecommand \translation [1]{[#1]}%
\providecommand \BibitemOpen [0]{}%
\providecommand \bibitemStop [0]{}%
\providecommand \bibitemNoStop [0]{.\EOS\space}%
\providecommand \EOS [0]{\spacefactor3000\relax}%
\providecommand \BibitemShut  [1]{\csname bibitem#1\endcsname}%
\let\auto@bib@innerbib\@empty
\bibitem [{\citenamefont {Vandersypen}\ \emph {et~al.}(2017)\citenamefont
  {Vandersypen}, \citenamefont {Bluhm}, \citenamefont {Clarke}, \citenamefont
  {Dzurak}, \citenamefont {Ishihara}, \citenamefont {Morello}, \citenamefont
  {Reilly}, \citenamefont {Schreiber},\ and\ \citenamefont
  {Veldhorst}}]{Vandersypen17-ISQ}%
  \BibitemOpen
  \bibfield  {author} {\bibinfo {author} {\bibfnamefont {L.~M.~K.}\
  \bibnamefont {Vandersypen}}, \bibinfo {author} {\bibfnamefont
  {H.}~\bibnamefont {Bluhm}}, \bibinfo {author} {\bibfnamefont {J.~S.}\
  \bibnamefont {Clarke}}, \bibinfo {author} {\bibfnamefont {A.~S.}\
  \bibnamefont {Dzurak}}, \bibinfo {author} {\bibfnamefont {R.}~\bibnamefont
  {Ishihara}}, \bibinfo {author} {\bibfnamefont {A.}~\bibnamefont {Morello}},
  \bibinfo {author} {\bibfnamefont {D.~J.}\ \bibnamefont {Reilly}}, \bibinfo
  {author} {\bibfnamefont {L.~R.}\ \bibnamefont {Schreiber}},\ and\ \bibinfo
  {author} {\bibfnamefont {M.}~\bibnamefont {Veldhorst}},\ }\bibfield  {title}
  {\bibinfo {title} {Interfacing spin qubits in quantum dots and donors—hot,
  dense, and coherent},\ }\href {https://doi.org/10.1038/s41534-017-0038-y}
  {\bibfield  {journal} {\bibinfo  {journal} {npj Quantum Inf.}\ }\textbf
  {\bibinfo {volume} {3}},\ \bibinfo {pages} {34} (\bibinfo {year}
  {2017})}\BibitemShut {NoStop}%
\bibitem [{\citenamefont {Li}\ \emph {et~al.}(2018)\citenamefont {Li},
  \citenamefont {Petit}, \citenamefont {Franke}, \citenamefont {Dehollain},
  \citenamefont {Helsen}, \citenamefont {Steudtner}, \citenamefont {Thomas},
  \citenamefont {Yoscovits}, \citenamefont {Singh}, \citenamefont {Wehner},
  \citenamefont {Vandersypen}, \citenamefont {Clarke},\ and\ \citenamefont
  {Veldhorst}}]{Li18-CNQ}%
  \BibitemOpen
  \bibfield  {author} {\bibinfo {author} {\bibfnamefont {R.}~\bibnamefont
  {Li}}, \bibinfo {author} {\bibfnamefont {L.}~\bibnamefont {Petit}}, \bibinfo
  {author} {\bibfnamefont {D.~P.}\ \bibnamefont {Franke}}, \bibinfo {author}
  {\bibfnamefont {J.~P.}\ \bibnamefont {Dehollain}}, \bibinfo {author}
  {\bibfnamefont {J.}~\bibnamefont {Helsen}}, \bibinfo {author} {\bibfnamefont
  {M.}~\bibnamefont {Steudtner}}, \bibinfo {author} {\bibfnamefont {N.~K.}\
  \bibnamefont {Thomas}}, \bibinfo {author} {\bibfnamefont {Z.~R.}\
  \bibnamefont {Yoscovits}}, \bibinfo {author} {\bibfnamefont {K.~J.}\
  \bibnamefont {Singh}}, \bibinfo {author} {\bibfnamefont {S.}~\bibnamefont
  {Wehner}}, \bibinfo {author} {\bibfnamefont {L.~M.~K.}\ \bibnamefont
  {Vandersypen}}, \bibinfo {author} {\bibfnamefont {J.~S.}\ \bibnamefont
  {Clarke}},\ and\ \bibinfo {author} {\bibfnamefont {M.}~\bibnamefont
  {Veldhorst}},\ }\bibfield  {title} {\bibinfo {title} {A crossbar network for
  silicon quantum dot qubits},\ }\href {https://doi.org/10.1126/sciadv.aar3960}
  {\bibfield  {journal} {\bibinfo  {journal} {Sci. Adv.}\ }\textbf {\bibinfo
  {volume} {4}},\ \bibinfo {pages} {eaar3960} (\bibinfo {year}
  {2018})}\BibitemShut {NoStop}%
\bibitem [{\citenamefont {Boter}\ \emph {et~al.}(2022)\citenamefont {Boter},
  \citenamefont {Dehollain}, \citenamefont {van Dijk}, \citenamefont {Xu},
  \citenamefont {Hensgens}, \citenamefont {Versluis}, \citenamefont {Naus},
  \citenamefont {Clarke}, \citenamefont {Veldhorst}, \citenamefont
  {Sebastiano},\ and\ \citenamefont {Vandersypen}}]{Boter22-SWA}%
  \BibitemOpen
  \bibfield  {author} {\bibinfo {author} {\bibfnamefont {J.~M.}\ \bibnamefont
  {Boter}}, \bibinfo {author} {\bibfnamefont {J.~P.}\ \bibnamefont
  {Dehollain}}, \bibinfo {author} {\bibfnamefont {J.~P.}\ \bibnamefont {van
  Dijk}}, \bibinfo {author} {\bibfnamefont {Y.}~\bibnamefont {Xu}}, \bibinfo
  {author} {\bibfnamefont {T.}~\bibnamefont {Hensgens}}, \bibinfo {author}
  {\bibfnamefont {R.}~\bibnamefont {Versluis}}, \bibinfo {author}
  {\bibfnamefont {H.~W.}\ \bibnamefont {Naus}}, \bibinfo {author}
  {\bibfnamefont {J.~S.}\ \bibnamefont {Clarke}}, \bibinfo {author}
  {\bibfnamefont {M.}~\bibnamefont {Veldhorst}}, \bibinfo {author}
  {\bibfnamefont {F.}~\bibnamefont {Sebastiano}},\ and\ \bibinfo {author}
  {\bibfnamefont {L.~M.}\ \bibnamefont {Vandersypen}},\ }\bibfield  {title}
  {\bibinfo {title} {Spiderweb array: A sparse spin-qubit array},\ }\href
  {https://doi.org/10.1103/PhysRevApplied.18.024053} {\bibfield  {journal}
  {\bibinfo  {journal} {Phys. Rev. Applied}\ }\textbf {\bibinfo {volume}
  {18}},\ \bibinfo {pages} {024053} (\bibinfo {year} {2022})}\BibitemShut
  {NoStop}%
\bibitem [{\citenamefont {Ansaloni}\ \emph {et~al.}(2020)\citenamefont
  {Ansaloni}, \citenamefont {Chatterjee}, \citenamefont {Bohuslavskyi},
  \citenamefont {Bertrand}, \citenamefont {Hutin}, \citenamefont {Vinet},\ and\
  \citenamefont {Kuemmeth}}]{Ansaloni20-FFQ}%
  \BibitemOpen
  \bibfield  {author} {\bibinfo {author} {\bibfnamefont {F.}~\bibnamefont
  {Ansaloni}}, \bibinfo {author} {\bibfnamefont {A.}~\bibnamefont
  {Chatterjee}}, \bibinfo {author} {\bibfnamefont {H.}~\bibnamefont
  {Bohuslavskyi}}, \bibinfo {author} {\bibfnamefont {B.}~\bibnamefont
  {Bertrand}}, \bibinfo {author} {\bibfnamefont {L.}~\bibnamefont {Hutin}},
  \bibinfo {author} {\bibfnamefont {M.}~\bibnamefont {Vinet}},\ and\ \bibinfo
  {author} {\bibfnamefont {F.}~\bibnamefont {Kuemmeth}},\ }\bibfield  {title}
  {\bibinfo {title} {Single-electron operations in a foundry-fabricated array
  of quantum dots},\ }\href {https://doi.org/10.1038/s41467-020-20280-3}
  {\bibfield  {journal} {\bibinfo  {journal} {Nat. Commun.}\ }\textbf {\bibinfo
  {volume} {11}},\ \bibinfo {pages} {6399} (\bibinfo {year}
  {2020})}\BibitemShut {NoStop}%
\bibitem [{\citenamefont {Pillarisetty}\ \emph {et~al.}(2021)\citenamefont
  {Pillarisetty}, \citenamefont {Watson}, \citenamefont {Mueller},
  \citenamefont {Henry}, \citenamefont {George}, \citenamefont {Bojarski},
  \citenamefont {Lampert}, \citenamefont {Luthi}, \citenamefont {Kotlyar},
  \citenamefont {Zietz}, \citenamefont {Neyens}, \citenamefont {Borjans},
  \citenamefont {Caudillo}, \citenamefont {Michalak}, \citenamefont {Nahm},
  \citenamefont {Park}, \citenamefont {Ramsey}, \citenamefont {Roberts},
  \citenamefont {Schaal}, \citenamefont {Zheng}, \citenamefont {Krähenmann},
  \citenamefont {Lodari}, \citenamefont {Zwerver}, \citenamefont {Veldhorst},
  \citenamefont {Scappucci}, \citenamefont {Vandersvpen},\ and\ \citenamefont
  {Clarke}}]{Pillarisetty21-SQC}%
  \BibitemOpen
  \bibfield  {author} {\bibinfo {author} {\bibfnamefont {R.}~\bibnamefont
  {Pillarisetty}}, \bibinfo {author} {\bibfnamefont {T.}~\bibnamefont
  {Watson}}, \bibinfo {author} {\bibfnamefont {B.}~\bibnamefont {Mueller}},
  \bibinfo {author} {\bibfnamefont {E.}~\bibnamefont {Henry}}, \bibinfo
  {author} {\bibfnamefont {H.}~\bibnamefont {George}}, \bibinfo {author}
  {\bibfnamefont {S.}~\bibnamefont {Bojarski}}, \bibinfo {author}
  {\bibfnamefont {L.}~\bibnamefont {Lampert}}, \bibinfo {author} {\bibfnamefont
  {F.}~\bibnamefont {Luthi}}, \bibinfo {author} {\bibfnamefont
  {R.}~\bibnamefont {Kotlyar}}, \bibinfo {author} {\bibfnamefont
  {O.}~\bibnamefont {Zietz}}, \bibinfo {author} {\bibfnamefont
  {S.}~\bibnamefont {Neyens}}, \bibinfo {author} {\bibfnamefont
  {F.}~\bibnamefont {Borjans}}, \bibinfo {author} {\bibfnamefont
  {R.}~\bibnamefont {Caudillo}}, \bibinfo {author} {\bibfnamefont
  {D.}~\bibnamefont {Michalak}}, \bibinfo {author} {\bibfnamefont
  {R.}~\bibnamefont {Nahm}}, \bibinfo {author} {\bibfnamefont {J.}~\bibnamefont
  {Park}}, \bibinfo {author} {\bibfnamefont {M.}~\bibnamefont {Ramsey}},
  \bibinfo {author} {\bibfnamefont {J.}~\bibnamefont {Roberts}}, \bibinfo
  {author} {\bibfnamefont {S.}~\bibnamefont {Schaal}}, \bibinfo {author}
  {\bibfnamefont {G.}~\bibnamefont {Zheng}}, \bibinfo {author} {\bibfnamefont
  {T.}~\bibnamefont {Krähenmann}}, \bibinfo {author} {\bibfnamefont
  {M.}~\bibnamefont {Lodari}}, \bibinfo {author} {\bibfnamefont
  {A.}~\bibnamefont {Zwerver}}, \bibinfo {author} {\bibfnamefont
  {M.}~\bibnamefont {Veldhorst}}, \bibinfo {author} {\bibfnamefont
  {G.}~\bibnamefont {Scappucci}}, \bibinfo {author} {\bibfnamefont
  {L.}~\bibnamefont {Vandersvpen}},\ and\ \bibinfo {author} {\bibfnamefont
  {J.}~\bibnamefont {Clarke}},\ }\bibfield  {title} {\bibinfo {title} {Si {MOS}
  and {Si/SiGe} quantum well spin qubit platforms for scalable quantum
  computing},\ }in\ \href {https://doi.org/10.1109/IEDM19574.2021.9720567}
  {\emph {\bibinfo {booktitle} {2021 IEEE Conference Proceedings}}},\ \bibinfo
  {series and number} {67th IEEE International Electron Devices Meeting
  (IEDM)}\ (\bibinfo  {publisher} {IEEE},\ \bibinfo {address} {San Francisco,
  CA, USA},\ \bibinfo {year} {2021})\ pp.\ \bibinfo {pages}
  {14.1.1--14.1.4}\BibitemShut {NoStop}%
\bibitem [{\citenamefont {Zwerver}\ \emph {et~al.}(2022)\citenamefont
  {Zwerver}, \citenamefont {Krähenmann}, \citenamefont {Watson}, \citenamefont
  {Lampert}, \citenamefont {George}, \citenamefont {Pillarisetty},
  \citenamefont {Bojarski}, \citenamefont {Amin}, \citenamefont {Amitonov},
  \citenamefont {Boter}, \citenamefont {Caudillo}, \citenamefont
  {Correas-Serrano}, \citenamefont {Dehollain}, \citenamefont {Droulers},
  \citenamefont {Henry}, \citenamefont {Kotlyar}, \citenamefont {L\"{u}thi},
  \citenamefont {Michalak}, \citenamefont {Mueller}, \citenamefont {Neyens},
  \citenamefont {Roberts}, \citenamefont {Samkharadze}, \citenamefont {Zheng},
  \citenamefont {Zietz}, \citenamefont {Scappucci}, \citenamefont {Veldhorst},
  \citenamefont {Vandersypen},\ and\ \citenamefont {Clarke}}]{Zwerver22-QMA}%
  \BibitemOpen
  \bibfield  {author} {\bibinfo {author} {\bibfnamefont {A.~M.~J.}\
  \bibnamefont {Zwerver}}, \bibinfo {author} {\bibfnamefont {T.}~\bibnamefont
  {Krähenmann}}, \bibinfo {author} {\bibfnamefont {T.~F.}\ \bibnamefont
  {Watson}}, \bibinfo {author} {\bibfnamefont {L.}~\bibnamefont {Lampert}},
  \bibinfo {author} {\bibfnamefont {H.~C.}\ \bibnamefont {George}}, \bibinfo
  {author} {\bibfnamefont {R.}~\bibnamefont {Pillarisetty}}, \bibinfo {author}
  {\bibfnamefont {S.~A.}\ \bibnamefont {Bojarski}}, \bibinfo {author}
  {\bibfnamefont {P.}~\bibnamefont {Amin}}, \bibinfo {author} {\bibfnamefont
  {S.~V.}\ \bibnamefont {Amitonov}}, \bibinfo {author} {\bibfnamefont {J.~M.}\
  \bibnamefont {Boter}}, \bibinfo {author} {\bibfnamefont {R.}~\bibnamefont
  {Caudillo}}, \bibinfo {author} {\bibfnamefont {D.}~\bibnamefont
  {Correas-Serrano}}, \bibinfo {author} {\bibfnamefont {J.~P.}\ \bibnamefont
  {Dehollain}}, \bibinfo {author} {\bibfnamefont {G.}~\bibnamefont {Droulers}},
  \bibinfo {author} {\bibfnamefont {E.~M.}\ \bibnamefont {Henry}}, \bibinfo
  {author} {\bibfnamefont {M.}~\bibnamefont {Kotlyar}, \bibfnamefont
  {R.vLodari}}, \bibinfo {author} {\bibfnamefont {F.}~\bibnamefont
  {L\"{u}thi}}, \bibinfo {author} {\bibfnamefont {D.~J.}\ \bibnamefont
  {Michalak}}, \bibinfo {author} {\bibfnamefont {B.~K.}\ \bibnamefont
  {Mueller}}, \bibinfo {author} {\bibfnamefont {S.}~\bibnamefont {Neyens}},
  \bibinfo {author} {\bibfnamefont {J.}~\bibnamefont {Roberts}}, \bibinfo
  {author} {\bibfnamefont {N.}~\bibnamefont {Samkharadze}}, \bibinfo {author}
  {\bibfnamefont {G.}~\bibnamefont {Zheng}}, \bibinfo {author} {\bibfnamefont
  {O.~K.}\ \bibnamefont {Zietz}}, \bibinfo {author} {\bibfnamefont
  {G.}~\bibnamefont {Scappucci}}, \bibinfo {author} {\bibfnamefont
  {M.}~\bibnamefont {Veldhorst}}, \bibinfo {author} {\bibfnamefont {L.~M.~K.}\
  \bibnamefont {Vandersypen}},\ and\ \bibinfo {author} {\bibfnamefont {J.~S.}\
  \bibnamefont {Clarke}},\ }\bibfield  {title} {\bibinfo {title} {Qubits made
  by advanced semiconductor manufacturing},\ }\href
  {https://doi.org/10.1038/s41928-022-00727-9} {\bibfield  {journal} {\bibinfo
  {journal} {Nat. Electron.}\ }\textbf {\bibinfo {volume} {5}},\ \bibinfo
  {pages} {184} (\bibinfo {year} {2022})}\BibitemShut {NoStop}%
\bibitem [{\citenamefont {Weinstein}\ \emph {et~al.}(2023)\citenamefont
  {Weinstein}, \citenamefont {Reed}, \citenamefont {Jones}, \citenamefont
  {Andrews}, \citenamefont {Barnes}, \citenamefont {Blumoff}, \citenamefont
  {Euliss}, \citenamefont {Eng}, \citenamefont {Fong}, \citenamefont {Ha},
  \citenamefont {Hulbert}, \citenamefont {Jackson}, \citenamefont {Jura},
  \citenamefont {Keating}, \citenamefont {Kerckhoff}, \citenamefont {Kiselev},
  \citenamefont {Matten}, \citenamefont {Sabbir}, \citenamefont {Smith},
  \citenamefont {Wright}, \citenamefont {Rakher}, \citenamefont {Ladd},\ and\
  \citenamefont {Borselli}}]{Weinstein22-ULS}%
  \BibitemOpen
  \bibfield  {author} {\bibinfo {author} {\bibfnamefont {A.~J.}\ \bibnamefont
  {Weinstein}}, \bibinfo {author} {\bibfnamefont {M.~D.}\ \bibnamefont {Reed}},
  \bibinfo {author} {\bibfnamefont {A.~M.}\ \bibnamefont {Jones}}, \bibinfo
  {author} {\bibfnamefont {R.~W.}\ \bibnamefont {Andrews}}, \bibinfo {author}
  {\bibfnamefont {D.}~\bibnamefont {Barnes}}, \bibinfo {author} {\bibfnamefont
  {J.~Z.}\ \bibnamefont {Blumoff}}, \bibinfo {author} {\bibfnamefont {L.~E.}\
  \bibnamefont {Euliss}}, \bibinfo {author} {\bibfnamefont {K.}~\bibnamefont
  {Eng}}, \bibinfo {author} {\bibfnamefont {B.}~\bibnamefont {Fong}}, \bibinfo
  {author} {\bibfnamefont {S.~D.}\ \bibnamefont {Ha}}, \bibinfo {author}
  {\bibfnamefont {D.~R.}\ \bibnamefont {Hulbert}}, \bibinfo {author}
  {\bibfnamefont {C.}~\bibnamefont {Jackson}}, \bibinfo {author} {\bibfnamefont
  {M.}~\bibnamefont {Jura}}, \bibinfo {author} {\bibfnamefont {T.~E.}\
  \bibnamefont {Keating}}, \bibinfo {author} {\bibfnamefont {J.}~\bibnamefont
  {Kerckhoff}}, \bibinfo {author} {\bibfnamefont {A.~A.}\ \bibnamefont
  {Kiselev}}, \bibinfo {author} {\bibfnamefont {J.}~\bibnamefont {Matten}},
  \bibinfo {author} {\bibfnamefont {G.}~\bibnamefont {Sabbir}}, \bibinfo
  {author} {\bibfnamefont {A.}~\bibnamefont {Smith}}, \bibinfo {author}
  {\bibfnamefont {J.}~\bibnamefont {Wright}}, \bibinfo {author} {\bibfnamefont
  {M.~T.}\ \bibnamefont {Rakher}}, \bibinfo {author} {\bibfnamefont {T.~D.}\
  \bibnamefont {Ladd}},\ and\ \bibinfo {author} {\bibfnamefont {M.~G.}\
  \bibnamefont {Borselli}},\ }\bibfield  {title} {\bibinfo {title} {Universal
  logic with encoded spin qubits in silicon},\ }\href
  {https://doi.org/10.1038/s41586-023-05777-3} {\bibfield  {journal} {\bibinfo
  {journal} {Nature}\ }\textbf {\bibinfo {volume} {615}},\ \bibinfo {pages}
  {817} (\bibinfo {year} {2023})}\BibitemShut {NoStop}%
\bibitem [{\citenamefont {McJunkin}(2021)}]{McJunkin21-PhD}%
  \BibitemOpen
  \bibfield  {author} {\bibinfo {author} {\bibfnamefont {T.~W.}\ \bibnamefont
  {McJunkin}},\ }\emph {\bibinfo {title} {Heterostructure Modifications,
  Fabrication Improvements, and Measurements Automation of {Si/SiGe} Quantum
  Dots for Quantum Computation}},\ \href
  {https://www.proquest.com/openview/bc78ff608faa683aee92fc66a414ad56/} {Ph.D.
  thesis},\ \bibinfo  {school} {The University of Wisconsin-Madison}, \bibinfo
  {address} {Madison, WI, USA} (\bibinfo {year} {2021})\BibitemShut {NoStop}%
\bibitem [{\citenamefont {Volk}\ \emph {et~al.}(2019)\citenamefont {Volk},
  \citenamefont {Zwerver}, \citenamefont {Mukhopadhyay}, \citenamefont
  {Eendebak}, \citenamefont {van Diepen}, \citenamefont {Dehollain},
  \citenamefont {Hensgens}, \citenamefont {Fujita}, \citenamefont {Reichl},
  \citenamefont {Wegscheider},\ and\ \citenamefont {Vandersypen}}]{Volk19-LQR}%
  \BibitemOpen
  \bibfield  {author} {\bibinfo {author} {\bibfnamefont {C.}~\bibnamefont
  {Volk}}, \bibinfo {author} {\bibfnamefont {A.~M.~J.}\ \bibnamefont
  {Zwerver}}, \bibinfo {author} {\bibfnamefont {U.}~\bibnamefont
  {Mukhopadhyay}}, \bibinfo {author} {\bibfnamefont {P.~T.}\ \bibnamefont
  {Eendebak}}, \bibinfo {author} {\bibfnamefont {C.~J.}\ \bibnamefont {van
  Diepen}}, \bibinfo {author} {\bibfnamefont {J.~P.}\ \bibnamefont
  {Dehollain}}, \bibinfo {author} {\bibfnamefont {T.}~\bibnamefont {Hensgens}},
  \bibinfo {author} {\bibfnamefont {T.}~\bibnamefont {Fujita}}, \bibinfo
  {author} {\bibfnamefont {C.}~\bibnamefont {Reichl}}, \bibinfo {author}
  {\bibfnamefont {W.}~\bibnamefont {Wegscheider}},\ and\ \bibinfo {author}
  {\bibfnamefont {L.~M.~K.}\ \bibnamefont {Vandersypen}},\ }\bibfield  {title}
  {\bibinfo {title} {Loading a quantum-dot based ``{Q}ubyte'' register},\
  }\href {https://doi.org/10.1038/s41534-019-0146-y} {\bibfield  {journal}
  {\bibinfo  {journal} {npj Quantum Inf.}\ }\textbf {\bibinfo {volume} {5}},\
  \bibinfo {pages} {29} (\bibinfo {year} {2019})}\BibitemShut {NoStop}%
\bibitem [{\citenamefont {Oosterkamp}\ \emph {et~al.}(1998)\citenamefont
  {Oosterkamp}, \citenamefont {Fujisawa}, \citenamefont {van~der Wiel},
  \citenamefont {Ishibashi}, \citenamefont {Hijman}, \citenamefont {Tarucha},\
  and\ \citenamefont {Kouwenhoven}}]{Oosterkamp98-MSQ}%
  \BibitemOpen
  \bibfield  {author} {\bibinfo {author} {\bibfnamefont {T.~H.}\ \bibnamefont
  {Oosterkamp}}, \bibinfo {author} {\bibfnamefont {T.}~\bibnamefont
  {Fujisawa}}, \bibinfo {author} {\bibfnamefont {W.~G.}\ \bibnamefont {van~der
  Wiel}}, \bibinfo {author} {\bibfnamefont {K.}~\bibnamefont {Ishibashi}},
  \bibinfo {author} {\bibfnamefont {R.~V.}\ \bibnamefont {Hijman}}, \bibinfo
  {author} {\bibfnamefont {S.}~\bibnamefont {Tarucha}},\ and\ \bibinfo {author}
  {\bibfnamefont {L.~P.}\ \bibnamefont {Kouwenhoven}},\ }\bibfield  {title}
  {\bibinfo {title} {Microwave spectroscopy of a quantum-dot molecule},\ }\href
  {https://doi.org/10.1038/27617} {\bibfield  {journal} {\bibinfo  {journal}
  {Nature}\ }\textbf {\bibinfo {volume} {395}},\ \bibinfo {pages} {873}
  (\bibinfo {year} {1998})}\BibitemShut {NoStop}%
\bibitem [{\citenamefont {Hensgens}\ \emph {et~al.}(2017)\citenamefont
  {Hensgens}, \citenamefont {Fujita}, \citenamefont {Janssen}, \citenamefont
  {Li}, \citenamefont {Van~Diepen}, \citenamefont {Reichl}, \citenamefont
  {Wegscheider}, \citenamefont {Sarma},\ and\ \citenamefont
  {Vandersypen}}]{Hensgens17-FHQ}%
  \BibitemOpen
  \bibfield  {author} {\bibinfo {author} {\bibfnamefont {T.}~\bibnamefont
  {Hensgens}}, \bibinfo {author} {\bibfnamefont {T.}~\bibnamefont {Fujita}},
  \bibinfo {author} {\bibfnamefont {L.}~\bibnamefont {Janssen}}, \bibinfo
  {author} {\bibfnamefont {X.}~\bibnamefont {Li}}, \bibinfo {author}
  {\bibfnamefont {C.~J.}\ \bibnamefont {Van~Diepen}}, \bibinfo {author}
  {\bibfnamefont {C.}~\bibnamefont {Reichl}}, \bibinfo {author} {\bibfnamefont
  {W.}~\bibnamefont {Wegscheider}}, \bibinfo {author} {\bibfnamefont {S.~D.}\
  \bibnamefont {Sarma}},\ and\ \bibinfo {author} {\bibfnamefont {L.~M.~K.}\
  \bibnamefont {Vandersypen}},\ }\bibfield  {title} {\bibinfo {title} {Quantum
  simulation of a fermi--hubbard model using a semiconductor quantum dot
  array},\ }\href {https://doi.org/doi.org/10.1038/nature23022} {\bibfield
  {journal} {\bibinfo  {journal} {Nature}\ }\textbf {\bibinfo {volume} {548}},\
  \bibinfo {pages} {70} (\bibinfo {year} {2017})}\BibitemShut {NoStop}%
\bibitem [{\citenamefont {Zwolak}\ and\ \citenamefont
  {Taylor}(2023)}]{Zwolak21-AAQ}%
  \BibitemOpen
  \bibfield  {author} {\bibinfo {author} {\bibfnamefont {J.~P.}\ \bibnamefont
  {Zwolak}}\ and\ \bibinfo {author} {\bibfnamefont {J.~M.}\ \bibnamefont
  {Taylor}},\ }\bibfield  {title} {\bibinfo {title} {{\it Colloquium}: Advances
  in automation of quantum dot devices control},\ }\href
  {https://doi.org/10.1103/RevModPhys.95.011006} {\bibfield  {journal}
  {\bibinfo  {journal} {Rev. Mod. Phys.}\ }\textbf {\bibinfo {volume} {95}},\
  \bibinfo {pages} {011006} (\bibinfo {year} {2023})}\BibitemShut {NoStop}%
\bibitem [{\citenamefont {Baart}\ \emph {et~al.}(2016)\citenamefont {Baart},
  \citenamefont {Eendebak}, \citenamefont {Reichl}, \citenamefont
  {Wegscheider},\ and\ \citenamefont {Vandersypen}}]{Baart16-CAT}%
  \BibitemOpen
  \bibfield  {author} {\bibinfo {author} {\bibfnamefont {T.~A.}\ \bibnamefont
  {Baart}}, \bibinfo {author} {\bibfnamefont {P.~T.}\ \bibnamefont {Eendebak}},
  \bibinfo {author} {\bibfnamefont {C.}~\bibnamefont {Reichl}}, \bibinfo
  {author} {\bibfnamefont {W.}~\bibnamefont {Wegscheider}},\ and\ \bibinfo
  {author} {\bibfnamefont {L.~M.~K.}\ \bibnamefont {Vandersypen}},\ }\bibfield
  {title} {\bibinfo {title} {Computer-automated tuning of semiconductor double
  quantum dots into the single-electron regime},\ }\href
  {https://doi.org/10.1063/1.4952624} {\bibfield  {journal} {\bibinfo
  {journal} {Appl. Phys. Lett.}\ }\textbf {\bibinfo {volume} {108}},\ \bibinfo
  {pages} {213104} (\bibinfo {year} {2016})}\BibitemShut {NoStop}%
\bibitem [{\citenamefont {Darulov\'a}\ \emph {et~al.}(2020)\citenamefont
  {Darulov\'a}, \citenamefont {Pauka}, \citenamefont {Wiebe}, \citenamefont
  {Chan}, \citenamefont {Gardener}, \citenamefont {Manfra}, \citenamefont
  {Cassidy},\ and\ \citenamefont {Troyer}}]{Darulova19-ATQ}%
  \BibitemOpen
  \bibfield  {author} {\bibinfo {author} {\bibfnamefont {J.}~\bibnamefont
  {Darulov\'a}}, \bibinfo {author} {\bibfnamefont {S.~J.}\ \bibnamefont
  {Pauka}}, \bibinfo {author} {\bibfnamefont {N.}~\bibnamefont {Wiebe}},
  \bibinfo {author} {\bibfnamefont {K.~W.}\ \bibnamefont {Chan}}, \bibinfo
  {author} {\bibfnamefont {G.~C.}\ \bibnamefont {Gardener}}, \bibinfo {author}
  {\bibfnamefont {M.~J.}\ \bibnamefont {Manfra}}, \bibinfo {author}
  {\bibfnamefont {M.~C.}\ \bibnamefont {Cassidy}},\ and\ \bibinfo {author}
  {\bibfnamefont {M.}~\bibnamefont {Troyer}},\ }\bibfield  {title} {\bibinfo
  {title} {Autonomous tuning and charge-state detection of gate-defined quantum
  dots},\ }\href {https://doi.org/10.1103/PhysRevApplied.13.054005} {\bibfield
  {journal} {\bibinfo  {journal} {Phys. Rev. Applied}\ }\textbf {\bibinfo
  {volume} {13}},\ \bibinfo {pages} {054005} (\bibinfo {year}
  {2020})}\BibitemShut {NoStop}%
\bibitem [{\citenamefont {Moon}\ \emph {et~al.}(2020)\citenamefont {Moon},
  \citenamefont {Lennon}, \citenamefont {Kirkpatrick}, \citenamefont {van
  Esbroeck}, \citenamefont {Camenzind}, \citenamefont {Yu}, \citenamefont
  {Vigneau}, \citenamefont {Zumb\"uhl}, \citenamefont {Briggs}, \citenamefont
  {Osborne}, \citenamefont {Sejdinovic}, \citenamefont {Laird},\ and\
  \citenamefont {Ares}}]{Moon20-ATQ}%
  \BibitemOpen
  \bibfield  {author} {\bibinfo {author} {\bibfnamefont {H.}~\bibnamefont
  {Moon}}, \bibinfo {author} {\bibfnamefont {D.~T.}\ \bibnamefont {Lennon}},
  \bibinfo {author} {\bibfnamefont {J.}~\bibnamefont {Kirkpatrick}}, \bibinfo
  {author} {\bibfnamefont {N.~M.}\ \bibnamefont {van Esbroeck}}, \bibinfo
  {author} {\bibfnamefont {L.~C.}\ \bibnamefont {Camenzind}}, \bibinfo {author}
  {\bibfnamefont {L.}~\bibnamefont {Yu}}, \bibinfo {author} {\bibfnamefont
  {F.}~\bibnamefont {Vigneau}}, \bibinfo {author} {\bibfnamefont {D.~M.}\
  \bibnamefont {Zumb\"uhl}}, \bibinfo {author} {\bibfnamefont {G.~A.~D.}\
  \bibnamefont {Briggs}}, \bibinfo {author} {\bibfnamefont {M.~A.}\
  \bibnamefont {Osborne}}, \bibinfo {author} {\bibfnamefont {D.}~\bibnamefont
  {Sejdinovic}}, \bibinfo {author} {\bibfnamefont {E.~A.}\ \bibnamefont
  {Laird}},\ and\ \bibinfo {author} {\bibfnamefont {N.}~\bibnamefont {Ares}},\
  }\bibfield  {title} {\bibinfo {title} {Machine learning enables completely
  automatic tuning of a quantum device faster than human experts},\ }\href
  {https://doi.org/10.1038/s41467-020-17835-9} {\bibfield  {journal} {\bibinfo
  {journal} {Nat. Commun.}\ }\textbf {\bibinfo {volume} {11}},\ \bibinfo
  {pages} {4161} (\bibinfo {year} {2020})}\BibitemShut {NoStop}%
\bibitem [{\citenamefont {Czischek}\ \emph {et~al.}(2022)\citenamefont
  {Czischek}, \citenamefont {Yon}, \citenamefont {Genest}, \citenamefont
  {Roux}, \citenamefont {Rochette}, \citenamefont {Lemyre}, \citenamefont
  {Moras}, \citenamefont {Pioro-Ladri\'ere}, \citenamefont {Drouin},
  \citenamefont {Beilliard},\ and\ \citenamefont {Melko}}]{Czischek21-MNA}%
  \BibitemOpen
  \bibfield  {author} {\bibinfo {author} {\bibfnamefont {S.}~\bibnamefont
  {Czischek}}, \bibinfo {author} {\bibfnamefont {V.}~\bibnamefont {Yon}},
  \bibinfo {author} {\bibfnamefont {M.-A.}\ \bibnamefont {Genest}}, \bibinfo
  {author} {\bibfnamefont {M.-A.}\ \bibnamefont {Roux}}, \bibinfo {author}
  {\bibfnamefont {S.}~\bibnamefont {Rochette}}, \bibinfo {author}
  {\bibfnamefont {J.~C.}\ \bibnamefont {Lemyre}}, \bibinfo {author}
  {\bibfnamefont {M.}~\bibnamefont {Moras}}, \bibinfo {author} {\bibfnamefont
  {M.}~\bibnamefont {Pioro-Ladri\'ere}}, \bibinfo {author} {\bibfnamefont
  {D.}~\bibnamefont {Drouin}}, \bibinfo {author} {\bibfnamefont
  {Y.}~\bibnamefont {Beilliard}},\ and\ \bibinfo {author} {\bibfnamefont
  {R.~G.}\ \bibnamefont {Melko}},\ }\bibfield  {title} {\bibinfo {title}
  {Miniaturizing neural networks for charge state autotuning in quantum dots},\
  }\href {https://doi.org/10.1088/2632-2153/ac34db} {\bibfield  {journal}
  {\bibinfo  {journal} {Mach. Learn.: Sci. Technol.}\ }\textbf {\bibinfo
  {volume} {3}},\ \bibinfo {pages} {015001} (\bibinfo {year}
  {2022})}\BibitemShut {NoStop}%
\bibitem [{\citenamefont {Zwolak}\ \emph {et~al.}(2018)\citenamefont {Zwolak},
  \citenamefont {Kalantre}, \citenamefont {Wu}, \citenamefont {Ragole},\ and\
  \citenamefont {Taylor}}]{Zwolak18-QLD}%
  \BibitemOpen
  \bibfield  {author} {\bibinfo {author} {\bibfnamefont {J.~P.}\ \bibnamefont
  {Zwolak}}, \bibinfo {author} {\bibfnamefont {S.~S.}\ \bibnamefont
  {Kalantre}}, \bibinfo {author} {\bibfnamefont {X.}~\bibnamefont {Wu}},
  \bibinfo {author} {\bibfnamefont {S.}~\bibnamefont {Ragole}},\ and\ \bibinfo
  {author} {\bibfnamefont {J.~M.}\ \bibnamefont {Taylor}},\ }\bibfield  {title}
  {\bibinfo {title} {{QFlow} lite dataset: {A} machine-learning approach to the
  charge states in quantum dot experiments},\ }\href
  {https://doi.org/10.1371/journal.pone.0205844} {\bibfield  {journal}
  {\bibinfo  {journal} {PLoS ONE}\ }\textbf {\bibinfo {volume} {13}},\ \bibinfo
  {pages} {e0205844} (\bibinfo {year} {2018})}\BibitemShut {NoStop}%
\bibitem [{\citenamefont {Durrer}\ \emph {et~al.}(2020)\citenamefont {Durrer},
  \citenamefont {Kratochwil}, \citenamefont {Koski}, \citenamefont {Landig},
  \citenamefont {Reichl}, \citenamefont {Wegscheider}, \citenamefont {Ihn},\
  and\ \citenamefont {Greplova}}]{Durrer19-ATQ}%
  \BibitemOpen
  \bibfield  {author} {\bibinfo {author} {\bibfnamefont {R.}~\bibnamefont
  {Durrer}}, \bibinfo {author} {\bibfnamefont {B.}~\bibnamefont {Kratochwil}},
  \bibinfo {author} {\bibfnamefont {J.}~\bibnamefont {Koski}}, \bibinfo
  {author} {\bibfnamefont {A.}~\bibnamefont {Landig}}, \bibinfo {author}
  {\bibfnamefont {C.}~\bibnamefont {Reichl}}, \bibinfo {author} {\bibfnamefont
  {W.}~\bibnamefont {Wegscheider}}, \bibinfo {author} {\bibfnamefont
  {T.}~\bibnamefont {Ihn}},\ and\ \bibinfo {author} {\bibfnamefont
  {E.}~\bibnamefont {Greplova}},\ }\bibfield  {title} {\bibinfo {title}
  {Automated tuning of double quantum dots into specific charge states using
  neural networks},\ }\href {https://doi.org/10.1103/PhysRevApplied.13.054019}
  {\bibfield  {journal} {\bibinfo  {journal} {Phys. Rev. Applied}\ }\textbf
  {\bibinfo {volume} {13}},\ \bibinfo {pages} {054019} (\bibinfo {year}
  {2020})}\BibitemShut {NoStop}%
\bibitem [{\citenamefont {Zwolak}\ \emph
  {et~al.}(2020{\natexlab{a}})\citenamefont {Zwolak}, \citenamefont {McJunkin},
  \citenamefont {Kalantre}, \citenamefont {Dodson}, \citenamefont {MacQuarrie},
  \citenamefont {Savage}, \citenamefont {Lagally}, \citenamefont {Coppersmith},
  \citenamefont {Eriksson},\ and\ \citenamefont {Taylor}}]{Zwolak20-AQD}%
  \BibitemOpen
  \bibfield  {author} {\bibinfo {author} {\bibfnamefont {J.~P.}\ \bibnamefont
  {Zwolak}}, \bibinfo {author} {\bibfnamefont {T.}~\bibnamefont {McJunkin}},
  \bibinfo {author} {\bibfnamefont {S.~S.}\ \bibnamefont {Kalantre}}, \bibinfo
  {author} {\bibfnamefont {J.}~\bibnamefont {Dodson}}, \bibinfo {author}
  {\bibfnamefont {E.~R.}\ \bibnamefont {MacQuarrie}}, \bibinfo {author}
  {\bibfnamefont {D.}~\bibnamefont {Savage}}, \bibinfo {author} {\bibfnamefont
  {M.}~\bibnamefont {Lagally}}, \bibinfo {author} {\bibfnamefont
  {S.}~\bibnamefont {Coppersmith}}, \bibinfo {author} {\bibfnamefont {M.~A.}\
  \bibnamefont {Eriksson}},\ and\ \bibinfo {author} {\bibfnamefont {J.~M.}\
  \bibnamefont {Taylor}},\ }\bibfield  {title} {\bibinfo {title} {Autotuning of
  double-dot devices in situ with machine learning},\ }\href
  {https://doi.org/10.1103/PhysRevApplied.13.034075} {\bibfield  {journal}
  {\bibinfo  {journal} {Phys. Rev. Applied}\ }\textbf {\bibinfo {volume}
  {13}},\ \bibinfo {pages} {034075} (\bibinfo {year}
  {2020}{\natexlab{a}})}\BibitemShut {NoStop}%
\bibitem [{\citenamefont {Zwolak}\ \emph {et~al.}(2021)\citenamefont {Zwolak},
  \citenamefont {McJunkin}, \citenamefont {Kalantre}, \citenamefont {Neyens},
  \citenamefont {MacQuarrie}, \citenamefont {Eriksson},\ and\ \citenamefont
  {Taylor}}]{Zwolak21-RBI}%
  \BibitemOpen
  \bibfield  {author} {\bibinfo {author} {\bibfnamefont {J.~P.}\ \bibnamefont
  {Zwolak}}, \bibinfo {author} {\bibfnamefont {T.}~\bibnamefont {McJunkin}},
  \bibinfo {author} {\bibfnamefont {S.~S.}\ \bibnamefont {Kalantre}}, \bibinfo
  {author} {\bibfnamefont {S.~F.}\ \bibnamefont {Neyens}}, \bibinfo {author}
  {\bibfnamefont {E.~R.}\ \bibnamefont {MacQuarrie}}, \bibinfo {author}
  {\bibfnamefont {M.~A.}\ \bibnamefont {Eriksson}},\ and\ \bibinfo {author}
  {\bibfnamefont {J.~M.}\ \bibnamefont {Taylor}},\ }\bibfield  {title}
  {\bibinfo {title} {Ray-based framework for state identification in quantum
  dot devices},\ }\href {https://doi.org/10.1103/PRXQuantum.2.020335}
  {\bibfield  {journal} {\bibinfo  {journal} {PRX Quantum}\ }\textbf {\bibinfo
  {volume} {2}},\ \bibinfo {pages} {020335} (\bibinfo {year}
  {2021})}\BibitemShut {NoStop}%
\bibitem [{\citenamefont {Lapointe-Major}\ \emph {et~al.}(2020)\citenamefont
  {Lapointe-Major}, \citenamefont {Germain}, \citenamefont {Camirand~Lemyre},
  \citenamefont {Lachance-Quirion}, \citenamefont {Rochette}, \citenamefont
  {Camirand~Lemyre},\ and\ \citenamefont
  {Pioro-Ladri\`ere}}]{Lapointe-Major19-ATQ}%
  \BibitemOpen
  \bibfield  {author} {\bibinfo {author} {\bibfnamefont {M.}~\bibnamefont
  {Lapointe-Major}}, \bibinfo {author} {\bibfnamefont {O.}~\bibnamefont
  {Germain}}, \bibinfo {author} {\bibfnamefont {J.}~\bibnamefont
  {Camirand~Lemyre}}, \bibinfo {author} {\bibfnamefont {D.}~\bibnamefont
  {Lachance-Quirion}}, \bibinfo {author} {\bibfnamefont {S.}~\bibnamefont
  {Rochette}}, \bibinfo {author} {\bibfnamefont {F.}~\bibnamefont
  {Camirand~Lemyre}},\ and\ \bibinfo {author} {\bibfnamefont {M.}~\bibnamefont
  {Pioro-Ladri\`ere}},\ }\bibfield  {title} {\bibinfo {title} {Algorithm for
  automated tuning of a quantum dot into the single-electron regime},\ }\href
  {https://doi.org/10.1103/PhysRevB.102.085301} {\bibfield  {journal} {\bibinfo
   {journal} {Phys. Rev. B}\ }\textbf {\bibinfo {volume} {102}},\ \bibinfo
  {pages} {085301} (\bibinfo {year} {2020})}\BibitemShut {NoStop}%
\bibitem [{\citenamefont {Darulov\'a}\ \emph {et~al.}(2021)\citenamefont
  {Darulov\'a}, \citenamefont {Troyer},\ and\ \citenamefont
  {Cassidy}}]{Darulova20-EDM}%
  \BibitemOpen
  \bibfield  {author} {\bibinfo {author} {\bibfnamefont {J.}~\bibnamefont
  {Darulov\'a}}, \bibinfo {author} {\bibfnamefont {M.}~\bibnamefont {Troyer}},\
  and\ \bibinfo {author} {\bibfnamefont {M.~C.}\ \bibnamefont {Cassidy}},\
  }\bibfield  {title} {\bibinfo {title} {Evaluation of synthetic and
  experimental training data in supervised machine learning applied to
  charge-state detection of quantum dots},\ }\href
  {https://doi.org/10.1088/2632-2153/ac104c} {\bibfield  {journal} {\bibinfo
  {journal} {Mach. Learn.: Sci. Technol.}\ }\textbf {\bibinfo {volume} {2}},\
  \bibinfo {pages} {045023} (\bibinfo {year} {2021})}\BibitemShut {NoStop}%
\bibitem [{\citenamefont {Ziegler}\ \emph {et~al.}(2022)\citenamefont
  {Ziegler}, \citenamefont {McJunkin}, \citenamefont {Joseph}, \citenamefont
  {Kalantre}, \citenamefont {Harpt}, \citenamefont {Savage}, \citenamefont
  {Lagally}, \citenamefont {Eriksson}, \citenamefont {Taylor},\ and\
  \citenamefont {Zwolak}}]{Ziegler22-TRA}%
  \BibitemOpen
  \bibfield  {author} {\bibinfo {author} {\bibfnamefont {J.}~\bibnamefont
  {Ziegler}}, \bibinfo {author} {\bibfnamefont {T.}~\bibnamefont {McJunkin}},
  \bibinfo {author} {\bibfnamefont {E.~S.}\ \bibnamefont {Joseph}}, \bibinfo
  {author} {\bibfnamefont {S.~S.}\ \bibnamefont {Kalantre}}, \bibinfo {author}
  {\bibfnamefont {B.}~\bibnamefont {Harpt}}, \bibinfo {author} {\bibfnamefont
  {D.~E.}\ \bibnamefont {Savage}}, \bibinfo {author} {\bibfnamefont {M.~G.}\
  \bibnamefont {Lagally}}, \bibinfo {author} {\bibfnamefont {M.~A.}\
  \bibnamefont {Eriksson}}, \bibinfo {author} {\bibfnamefont {J.~M.}\
  \bibnamefont {Taylor}},\ and\ \bibinfo {author} {\bibfnamefont {J.~P.}\
  \bibnamefont {Zwolak}},\ }\bibfield  {title} {\bibinfo {title} {Toward robust
  autotuning of noisy quantum dot devices},\ }\href
  {https://doi.org/10.1103/PhysRevApplied.17.024069} {\bibfield  {journal}
  {\bibinfo  {journal} {Phys. Rev. Applied}\ }\textbf {\bibinfo {volume}
  {17}},\ \bibinfo {pages} {024069} (\bibinfo {year} {2022})}\BibitemShut
  {NoStop}%
\bibitem [{\citenamefont {Zwolak}\ \emph {et~al.}(2022)\citenamefont {Zwolak},
  \citenamefont {Taylor}, \citenamefont {Kalantre},\ and\ \citenamefont
  {McJunkin}}]{Zwolak22-Patent}%
  \BibitemOpen
  \bibfield  {author} {\bibinfo {author} {\bibfnamefont {J.~P.}\ \bibnamefont
  {Zwolak}}, \bibinfo {author} {\bibfnamefont {J.~M.}\ \bibnamefont {Taylor}},
  \bibinfo {author} {\bibfnamefont {S.~S.}\ \bibnamefont {Kalantre}},\ and\
  \bibinfo {author} {\bibfnamefont {T.~W.}\ \bibnamefont {McJunkin}},\
  }\href@noop {} {\bibinfo {title} {Ray-based classifier apparatus and tuning a
  device using machine learning with a ray-based classification framework}},\
  \bibinfo {howpublished} {WIPO IP Portal} (\bibinfo {year} {2022})\BibitemShut
  {NoStop}%
\bibitem [{\citenamefont {Hader}\ \emph {et~al.}(2023)\citenamefont {Hader},
  \citenamefont {Vogelbruch}, \citenamefont {Humpohl}, \citenamefont
  {Hangleiter}, \citenamefont {Eguzo}, \citenamefont {Heinen}, \citenamefont
  {Meyer},\ and\ \citenamefont {van Waasen}}]{Hader23-NST}%
  \BibitemOpen
  \bibfield  {author} {\bibinfo {author} {\bibfnamefont {F.}~\bibnamefont
  {Hader}}, \bibinfo {author} {\bibfnamefont {J.}~\bibnamefont {Vogelbruch}},
  \bibinfo {author} {\bibfnamefont {S.}~\bibnamefont {Humpohl}}, \bibinfo
  {author} {\bibfnamefont {T.}~\bibnamefont {Hangleiter}}, \bibinfo {author}
  {\bibfnamefont {C.}~\bibnamefont {Eguzo}}, \bibinfo {author} {\bibfnamefont
  {S.}~\bibnamefont {Heinen}}, \bibinfo {author} {\bibfnamefont
  {S.}~\bibnamefont {Meyer}},\ and\ \bibinfo {author} {\bibfnamefont
  {S.}~\bibnamefont {van Waasen}},\ }\bibfield  {title} {\bibinfo {title} {On
  noise-sensitive automatic tuning of gate-defined sensor dots},\ }\href
  {https://doi.org/10.1109/TQE.2023.3255743} {\bibfield  {journal} {\bibinfo
  {journal} {IEEE Trans. Quantum Eng.}\ }\textbf {\bibinfo {volume} {4}},\
  \bibinfo {pages} {5500218} (\bibinfo {year} {2023})}\BibitemShut {NoStop}%
\bibitem [{\citenamefont {{National Institute of Standards and
  Technology}}(2022)}]{qf-data}%
  \BibitemOpen
  \bibfield  {author} {\bibinfo {author} {\bibnamefont {{National Institute of
  Standards and Technology}}},\ }\href@noop {} {\bibinfo {title} {Qflow 2.0:
  Quantum dot data for machine learning}},\ \bibinfo {howpublished} {Database:
  data.nist.gov, \url{https://doi.org/10.18434/T4/1423788}} (\bibinfo {year}
  {2022})\BibitemShut {NoStop}%
\bibitem [{\citenamefont {Liu}\ \emph {et~al.}(2022)\citenamefont {Liu},
  \citenamefont {Wang}, \citenamefont {Wang}, \citenamefont {Sun},
  \citenamefont {Yin}, \citenamefont {Li}, \citenamefont {Cao},\ and\
  \citenamefont {Guo}}]{Liu22-ACT}%
  \BibitemOpen
  \bibfield  {author} {\bibinfo {author} {\bibfnamefont {H.}~\bibnamefont
  {Liu}}, \bibinfo {author} {\bibfnamefont {B.}~\bibnamefont {Wang}}, \bibinfo
  {author} {\bibfnamefont {N.}~\bibnamefont {Wang}}, \bibinfo {author}
  {\bibfnamefont {Z.}~\bibnamefont {Sun}}, \bibinfo {author} {\bibfnamefont
  {H.}~\bibnamefont {Yin}}, \bibinfo {author} {\bibfnamefont {H.}~\bibnamefont
  {Li}}, \bibinfo {author} {\bibfnamefont {G.}~\bibnamefont {Cao}},\ and\
  \bibinfo {author} {\bibfnamefont {G.}~\bibnamefont {Guo}},\ }\bibfield
  {title} {\bibinfo {title} {An automated approach for consecutive tuning of
  quantum dot arrays},\ }\href {https://doi.org/10.1063/5.0111128} {\bibfield
  {journal} {\bibinfo  {journal} {Appl. Phys. Lett.}\ }\textbf {\bibinfo
  {volume} {121}},\ \bibinfo {pages} {084002} (\bibinfo {year}
  {2022})}\BibitemShut {NoStop}%
\bibitem [{\citenamefont {Kalantre}\ \emph {et~al.}(2019)\citenamefont
  {Kalantre}, \citenamefont {Zwolak}, \citenamefont {Ragole}, \citenamefont
  {Wu}, \citenamefont {Zimmerman}, \citenamefont {Stewart},\ and\ \citenamefont
  {Taylor}}]{Kalantre17-MLD}%
  \BibitemOpen
  \bibfield  {author} {\bibinfo {author} {\bibfnamefont {S.~S.}\ \bibnamefont
  {Kalantre}}, \bibinfo {author} {\bibfnamefont {J.~P.}\ \bibnamefont
  {Zwolak}}, \bibinfo {author} {\bibfnamefont {S.}~\bibnamefont {Ragole}},
  \bibinfo {author} {\bibfnamefont {X.}~\bibnamefont {Wu}}, \bibinfo {author}
  {\bibfnamefont {N.~M.}\ \bibnamefont {Zimmerman}}, \bibinfo {author}
  {\bibfnamefont {M.~D.}\ \bibnamefont {Stewart}},\ and\ \bibinfo {author}
  {\bibfnamefont {J.~M.}\ \bibnamefont {Taylor}},\ }\bibfield  {title}
  {\bibinfo {title} {Machine learning techniques for state recognition and
  auto-tuning in quantum dots},\ }\href
  {https://doi.org/10.1038/s41534-018-0118-7} {\bibfield  {journal} {\bibinfo
  {journal} {npj Quantum Inf.}\ }\textbf {\bibinfo {volume} {5}},\ \bibinfo
  {pages} {1} (\bibinfo {year} {2019})}\BibitemShut {NoStop}%
\bibitem [{\citenamefont {Zwolak}\ \emph
  {et~al.}(2020{\natexlab{b}})\citenamefont {Zwolak}, \citenamefont {Kalantre},
  \citenamefont {McJunkin}, \citenamefont {Weber},\ and\ \citenamefont
  {Taylor}}]{Zwolak20-RBC}%
  \BibitemOpen
  \bibfield  {author} {\bibinfo {author} {\bibfnamefont {J.~P.}\ \bibnamefont
  {Zwolak}}, \bibinfo {author} {\bibfnamefont {S.~S.}\ \bibnamefont
  {Kalantre}}, \bibinfo {author} {\bibfnamefont {T.}~\bibnamefont {McJunkin}},
  \bibinfo {author} {\bibfnamefont {B.~J.}\ \bibnamefont {Weber}},\ and\
  \bibinfo {author} {\bibfnamefont {J.~M.}\ \bibnamefont {Taylor}},\ }\bibfield
   {title} {\bibinfo {title} {Ray-based classification framework for
  high-dimensional data},\ }in\ \href@noop {} {\emph {\bibinfo {booktitle}
  {Third Workshop on Machine Learning and the Physical Sciences (NeurIPS
  2020)}}}\ (\bibinfo {address} {Vancouver, Canada},\ \bibinfo {year} {2020})\
  pp.\ \bibinfo {pages} {1--7},\ \bibinfo {note} {arXiv:2010.00500}\BibitemShut
  {NoStop}%
\bibitem [{\citenamefont {Chatterjee}\ \emph {et~al.}(2022)\citenamefont
  {Chatterjee}, \citenamefont {Ansaloni}, \citenamefont {Rasmussen},
  \citenamefont {Brovang}, \citenamefont {Fedele}, \citenamefont
  {Bohuslavskyi}, \citenamefont {Krause},\ and\ \citenamefont
  {Kuemmeth}}]{Chatterjee21-AEC}%
  \BibitemOpen
  \bibfield  {author} {\bibinfo {author} {\bibfnamefont {A.}~\bibnamefont
  {Chatterjee}}, \bibinfo {author} {\bibfnamefont {F.}~\bibnamefont
  {Ansaloni}}, \bibinfo {author} {\bibfnamefont {T.}~\bibnamefont {Rasmussen}},
  \bibinfo {author} {\bibfnamefont {B.}~\bibnamefont {Brovang}}, \bibinfo
  {author} {\bibfnamefont {F.}~\bibnamefont {Fedele}}, \bibinfo {author}
  {\bibfnamefont {H.}~\bibnamefont {Bohuslavskyi}}, \bibinfo {author}
  {\bibfnamefont {O.}~\bibnamefont {Krause}},\ and\ \bibinfo {author}
  {\bibfnamefont {F.}~\bibnamefont {Kuemmeth}},\ }\bibfield  {title} {\bibinfo
  {title} {Autonomous estimation of high-dimensional coulomb diamonds from
  sparse measurements},\ }\href
  {https://doi.org/10.1103/PhysRevApplied.18.064040} {\bibfield  {journal}
  {\bibinfo  {journal} {Phys. Rev. Applied}\ }\textbf {\bibinfo {volume}
  {18}},\ \bibinfo {pages} {064040} (\bibinfo {year} {2022})}\BibitemShut
  {NoStop}%
\bibitem [{\citenamefont {Hensgens}(2018)}]{Hensgens18-PhD}%
  \BibitemOpen
  \bibfield  {author} {\bibinfo {author} {\bibfnamefont {T.}~\bibnamefont
  {Hensgens}},\ }\emph {\bibinfo {title} {Emulating Fermi-Hubbard physics with
  quantum dots: from few to more and how to}},\ \href
  {https://doi.org/10.4233/uuid:b71f3b0b-73a0-4996-896c-84ed43e72035} {Ph.D.
  thesis},\ \bibinfo  {school} {Delft University of Technology}, \bibinfo
  {address} {Delft, Netherlands} (\bibinfo {year} {2018})\BibitemShut {NoStop}%
\bibitem [{\citenamefont {Perron}\ \emph {et~al.}(2015)\citenamefont {Perron},
  \citenamefont {Stewart~Jr},\ and\ \citenamefont {Zimmerman}}]{Perron15-QSB}%
  \BibitemOpen
  \bibfield  {author} {\bibinfo {author} {\bibfnamefont {J.~K.}\ \bibnamefont
  {Perron}}, \bibinfo {author} {\bibfnamefont {M.~D.}\ \bibnamefont
  {Stewart~Jr}},\ and\ \bibinfo {author} {\bibfnamefont {N.~M.}\ \bibnamefont
  {Zimmerman}},\ }\bibfield  {title} {\bibinfo {title} {A quantitative study of
  bias triangles presented in chemical potential space},\ }\href
  {https://doi.org/10.1088/0953-8984/27/23/235302} {\bibfield  {journal}
  {\bibinfo  {journal} {J. Phys.: Condens. Matter}\ }\textbf {\bibinfo {volume}
  {27}},\ \bibinfo {pages} {235302} (\bibinfo {year} {2015})}\BibitemShut
  {NoStop}%
\bibitem [{\citenamefont {Virtanen}\ \emph {et~al.}(2020)\citenamefont
  {Virtanen}, \citenamefont {Gommers}, \citenamefont {Oliphant}, \citenamefont
  {Haberland}, \citenamefont {Reddy}, \citenamefont {Cournapeau}, \citenamefont
  {Burovski}, \citenamefont {Peterson}, \citenamefont {Weckesser},
  \citenamefont {Bright}, \citenamefont {{van der Walt}}, \citenamefont
  {Brett}, \citenamefont {Wilson}, \citenamefont {Millman}, \citenamefont
  {Mayorov}, \citenamefont {Nelson}, \citenamefont {Jones}, \citenamefont
  {Kern}, \citenamefont {Larson}, \citenamefont {Carey}, \citenamefont {Polat},
  \citenamefont {Feng}, \citenamefont {Moore}, \citenamefont {{VanderPlas}},
  \citenamefont {Laxalde}, \citenamefont {Perktold}, \citenamefont {Cimrman},
  \citenamefont {Henriksen}, \citenamefont {Quintero}, \citenamefont {Harris},
  \citenamefont {Archibald}, \citenamefont {Ribeiro}, \citenamefont
  {Pedregosa}, \citenamefont {{van Mulbregt}},\ and\ \citenamefont {{SciPy 1.0
  Contributors}}}]{SciPy}%
  \BibitemOpen
  \bibfield  {author} {\bibinfo {author} {\bibfnamefont {P.}~\bibnamefont
  {Virtanen}}, \bibinfo {author} {\bibfnamefont {R.}~\bibnamefont {Gommers}},
  \bibinfo {author} {\bibfnamefont {T.~E.}\ \bibnamefont {Oliphant}}, \bibinfo
  {author} {\bibfnamefont {M.}~\bibnamefont {Haberland}}, \bibinfo {author}
  {\bibfnamefont {T.}~\bibnamefont {Reddy}}, \bibinfo {author} {\bibfnamefont
  {D.}~\bibnamefont {Cournapeau}}, \bibinfo {author} {\bibfnamefont
  {E.}~\bibnamefont {Burovski}}, \bibinfo {author} {\bibfnamefont
  {P.}~\bibnamefont {Peterson}}, \bibinfo {author} {\bibfnamefont
  {W.}~\bibnamefont {Weckesser}}, \bibinfo {author} {\bibfnamefont
  {J.}~\bibnamefont {Bright}}, \bibinfo {author} {\bibfnamefont {S.~J.}\
  \bibnamefont {{van der Walt}}}, \bibinfo {author} {\bibfnamefont
  {M.}~\bibnamefont {Brett}}, \bibinfo {author} {\bibfnamefont
  {J.}~\bibnamefont {Wilson}}, \bibinfo {author} {\bibfnamefont {K.~J.}\
  \bibnamefont {Millman}}, \bibinfo {author} {\bibfnamefont {N.}~\bibnamefont
  {Mayorov}}, \bibinfo {author} {\bibfnamefont {A.~R.~J.}\ \bibnamefont
  {Nelson}}, \bibinfo {author} {\bibfnamefont {E.}~\bibnamefont {Jones}},
  \bibinfo {author} {\bibfnamefont {R.}~\bibnamefont {Kern}}, \bibinfo {author}
  {\bibfnamefont {E.}~\bibnamefont {Larson}}, \bibinfo {author} {\bibfnamefont
  {C.~J.}\ \bibnamefont {Carey}}, \bibinfo {author} {\bibfnamefont
  {{\.I}.}~\bibnamefont {Polat}}, \bibinfo {author} {\bibfnamefont
  {Y.}~\bibnamefont {Feng}}, \bibinfo {author} {\bibfnamefont {E.~W.}\
  \bibnamefont {Moore}}, \bibinfo {author} {\bibfnamefont {J.}~\bibnamefont
  {{VanderPlas}}}, \bibinfo {author} {\bibfnamefont {D.}~\bibnamefont
  {Laxalde}}, \bibinfo {author} {\bibfnamefont {J.}~\bibnamefont {Perktold}},
  \bibinfo {author} {\bibfnamefont {R.}~\bibnamefont {Cimrman}}, \bibinfo
  {author} {\bibfnamefont {I.}~\bibnamefont {Henriksen}}, \bibinfo {author}
  {\bibfnamefont {E.~A.}\ \bibnamefont {Quintero}}, \bibinfo {author}
  {\bibfnamefont {C.~R.}\ \bibnamefont {Harris}}, \bibinfo {author}
  {\bibfnamefont {A.~M.}\ \bibnamefont {Archibald}}, \bibinfo {author}
  {\bibfnamefont {A.~H.}\ \bibnamefont {Ribeiro}}, \bibinfo {author}
  {\bibfnamefont {F.}~\bibnamefont {Pedregosa}}, \bibinfo {author}
  {\bibfnamefont {P.}~\bibnamefont {{van Mulbregt}}},\ and\ \bibinfo {author}
  {\bibnamefont {{SciPy 1.0 Contributors}}},\ }\bibfield  {title} {\bibinfo
  {title} {{{SciPy} 1.0: Fundamental Algorithms for Scientific Computing in
  Python}},\ }\href {https://doi.org/10.1038/s41592-019-0686-2} {\bibfield
  {journal} {\bibinfo  {journal} {Nat. Methods}\ }\textbf {\bibinfo {volume}
  {17}},\ \bibinfo {pages} {261} (\bibinfo {year} {2020})}\BibitemShut
  {NoStop}%
\bibitem [{Note1()}]{Note1}%
  \BibitemOpen
  \bibinfo {note} {Both methods are implemented in the current version of the
  PIT algorithm.}\BibitemShut {Stop}%
\bibitem [{\citenamefont {Lin}\ \emph {et~al.}(2016)\citenamefont {Lin},
  \citenamefont {Dollár}, \citenamefont {Girshick}, \citenamefont {He},
  \citenamefont {Hariharan},\ and\ \citenamefont {Belongie}}]{Lin16-FPN}%
  \BibitemOpen
  \bibfield  {author} {\bibinfo {author} {\bibfnamefont {T.-Y.}\ \bibnamefont
  {Lin}}, \bibinfo {author} {\bibfnamefont {P.}~\bibnamefont {Dollár}},
  \bibinfo {author} {\bibfnamefont {R.}~\bibnamefont {Girshick}}, \bibinfo
  {author} {\bibfnamefont {K.}~\bibnamefont {He}}, \bibinfo {author}
  {\bibfnamefont {B.}~\bibnamefont {Hariharan}},\ and\ \bibinfo {author}
  {\bibfnamefont {S.}~\bibnamefont {Belongie}},\ }\bibfield  {title} {\bibinfo
  {title} {Feature pyramid networks for object detection},\ }\href
  {https://arxiv.org/abs/1612.03144} {\bibfield  {journal} {\bibinfo  {journal}
  {arXiv:1612.03144}\ } (\bibinfo {year} {2016})}\BibitemShut {NoStop}%
\bibitem [{\citenamefont {Ziegler}\ \emph {et~al.}(2023)\citenamefont
  {Ziegler}, \citenamefont {Luthi}, \citenamefont {Ramsey}, \citenamefont
  {Borjans}, \citenamefont {Zheng},\ and\ \citenamefont
  {Zwolak}}]{Ziegler23-AEC}%
  \BibitemOpen
  \bibfield  {author} {\bibinfo {author} {\bibfnamefont {J.}~\bibnamefont
  {Ziegler}}, \bibinfo {author} {\bibfnamefont {F.}~\bibnamefont {Luthi}},
  \bibinfo {author} {\bibfnamefont {M.}~\bibnamefont {Ramsey}}, \bibinfo
  {author} {\bibfnamefont {F.}~\bibnamefont {Borjans}}, \bibinfo {author}
  {\bibfnamefont {G.}~\bibnamefont {Zheng}},\ and\ \bibinfo {author}
  {\bibfnamefont {J.~P.}\ \bibnamefont {Zwolak}},\ }\bibfield  {title}
  {\bibinfo {title} {Automated extraction of capacitive coupling for quantum
  dot systems},\ }\href {https://doi.org/10.1103/PhysRevApplied.19.054077}
  {\bibfield  {journal} {\bibinfo  {journal} {Phys. Rev. Applied}\ }\textbf
  {\bibinfo {volume} {19}},\ \bibinfo {pages} {054077} (\bibinfo {year}
  {2023})}\BibitemShut {NoStop}%
\bibitem [{\citenamefont {Mills}\ \emph {et~al.}(2019)\citenamefont {Mills},
  \citenamefont {Zajac}, \citenamefont {Gullans}, \citenamefont {Schupp},
  \citenamefont {Hazard},\ and\ \citenamefont {Petta}}]{Mills19-SSC}%
  \BibitemOpen
  \bibfield  {author} {\bibinfo {author} {\bibfnamefont {A.~R.}\ \bibnamefont
  {Mills}}, \bibinfo {author} {\bibfnamefont {D.~M.}\ \bibnamefont {Zajac}},
  \bibinfo {author} {\bibfnamefont {M.~J.}\ \bibnamefont {Gullans}}, \bibinfo
  {author} {\bibfnamefont {F.~J.}\ \bibnamefont {Schupp}}, \bibinfo {author}
  {\bibfnamefont {T.~M.}\ \bibnamefont {Hazard}},\ and\ \bibinfo {author}
  {\bibfnamefont {J.~R.}\ \bibnamefont {Petta}},\ }\bibfield  {title} {\bibinfo
  {title} {Shuttling a single charge across a one-dimensional array of silicon
  quantum dots},\ }\href {https://doi.org/10.1038/s41467-019-08970-z}
  {\bibfield  {journal} {\bibinfo  {journal} {Nat. Commun.}\ }\textbf {\bibinfo
  {volume} {10}},\ \bibinfo {pages} {1063} (\bibinfo {year}
  {2019})}\BibitemShut {NoStop}%
\bibitem [{Note2()}]{Note2}%
  \BibitemOpen
  \bibinfo {note} {The $20~\%$ of ray length requirement is determined by the
  lower bound on the distance from the peak necessary to properly identify
  it.}\BibitemShut {Stop}%
\bibitem [{Note3()}]{Note3}%
  \BibitemOpen
  \bibinfo {note} {The noise distribution extracted from 756 small experimental
  scans ranges from $0.05$ to $5.35$ in the units of the optimized noise
  configuration from Ref.~\cite {Ziegler22-TRA}, with the noise level mean $\mu
  =0.3(4)$ and median $M=0.2$ (median absolute deviation MAD=0.07). Since the
  data is highly skewed, with the Fisher-Pearson coefficient of skewness
  $g_1=9.1$, we opt to use the median as a measure of central tendency for the
  noise distribution.}\BibitemShut {Stop}%
\bibitem [{\citenamefont {Hastings}(1970)}]{Hasting70-MCS}%
  \BibitemOpen
  \bibfield  {author} {\bibinfo {author} {\bibfnamefont {W.~K.}\ \bibnamefont
  {Hastings}},\ }\bibfield  {title} {\bibinfo {title} {{Monte Carlo sampling
  methods using Markov chains and their applications}},\ }\href
  {https://doi.org/10.1093/biomet/57.1.97} {\bibfield  {journal} {\bibinfo
  {journal} {Biometrika}\ }\textbf {\bibinfo {volume} {57}},\ \bibinfo {pages}
  {97} (\bibinfo {year} {1970})},\ \Eprint
  {https://arxiv.org/abs/https://academic.oup.com/biomet/article-pdf/57/1/97/23940249/57-1-97.pdf}
  {https://academic.oup.com/biomet/article-pdf/57/1/97/23940249/57-1-97.pdf}
  \BibitemShut {NoStop}%
\bibitem [{Note4()}]{Note4}%
  \BibitemOpen
  \bibinfo {note} {We use a notation value(uncertainty) to express
  uncertainties, for example, $1.5(6)~\si {\centi \meter }$ would be
  interpreted as $(1.5\pm 0.6)~\si {\centi \meter }$. All uncertainties herein
  reflect the uncorrelated combination of single-standard deviation statistical
  and systematic uncertainties.}\BibitemShut {Stop}%
\bibitem [{\citenamefont {Nelder}\ and\ \citenamefont
  {Mead}(1965)}]{Nelder65-NMA}%
  \BibitemOpen
  \bibfield  {author} {\bibinfo {author} {\bibfnamefont {J.~A.}\ \bibnamefont
  {Nelder}}\ and\ \bibinfo {author} {\bibfnamefont {R.}~\bibnamefont {Mead}},\
  }\bibfield  {title} {\bibinfo {title} {A simplex method for function
  minimization},\ }\href {https://doi.org/10.1093/comjnl/7.4.308} {\bibfield
  {journal} {\bibinfo  {journal} {Comput. J.}\ }\textbf {\bibinfo {volume}
  {7}},\ \bibinfo {pages} {308} (\bibinfo {year} {1965})}\BibitemShut {NoStop}%
\bibitem [{\citenamefont {Gao}\ and\ \citenamefont {Han}(2012)}]{Gao12-IMN}%
  \BibitemOpen
  \bibfield  {author} {\bibinfo {author} {\bibfnamefont {F.}~\bibnamefont
  {Gao}}\ and\ \bibinfo {author} {\bibfnamefont {L.}~\bibnamefont {Han}},\
  }\bibfield  {title} {\bibinfo {title} {Implementing the nelder-mead simplex
  algorithm with adaptive parameters},\ }\href
  {https://doi.org/10.1007/s10589-010-9329-3} {\bibfield  {journal} {\bibinfo
  {journal} {Comput. Optim. Appl.}\ }\textbf {\bibinfo {volume} {51}},\
  \bibinfo {pages} {259} (\bibinfo {year} {2012})}\BibitemShut {NoStop}%
\bibitem [{\citenamefont {Caticha}(2022)}]{Caticha22-Ent}%
  \BibitemOpen
  \bibfield  {author} {\bibinfo {author} {\bibfnamefont {A.}~\bibnamefont
  {Caticha}},\ }\bibfield  {title} {\bibinfo {title} {Entropic physics:
  Probability, entropy, and the foundations of physics},\ }\href {https://www.
  albany. edu/physics/faculty/ariel-caticha} {\bibfield  {journal} {\bibinfo
  {journal} {Online at https://www. albany. edu/physics/faculty/ariel-caticha}\
  } (\bibinfo {year} {2022})},\ \bibinfo {note} {{A}ccessed on 30 August
  2022}\BibitemShut {NoStop}%
\bibitem [{\citenamefont {Pessoa}\ \emph {et~al.}(2021)\citenamefont {Pessoa},
  \citenamefont {Costa},\ and\ \citenamefont {Caticha}}]{Caticha21-Gibbs}%
  \BibitemOpen
  \bibfield  {author} {\bibinfo {author} {\bibfnamefont {P.}~\bibnamefont
  {Pessoa}}, \bibinfo {author} {\bibfnamefont {F.~X.}\ \bibnamefont {Costa}},\
  and\ \bibinfo {author} {\bibfnamefont {A.}~\bibnamefont {Caticha}},\
  }\bibfield  {title} {\bibinfo {title} {Entropic dynamics on gibbs statistical
  manifolds},\ }\bibfield  {journal} {\bibinfo  {journal} {Entropy}\ }\textbf
  {\bibinfo {volume} {23}},\ \href {https://doi.org/10.3390/e23050494}
  {10.3390/e23050494} (\bibinfo {year} {2021})\BibitemShut {NoStop}%
\end{thebibliography}
\end{document}